\documentclass{ws-ijbc}
\usepackage{ws-rotating}     % used only when sideways tables/figures are used
\begin{document}

\catchline{}{}{}{}{} % Publisher's Area please ignore

\markboth{Constantino Tsallis}{Some open points in nonextensive statistical mechanics}

\title{ (This paper is for the Special Issue edited by \\ Prof. Gregoire Nicolis, Prof. Marko Robnik, Dr. Vassilis Rothos and Dr. Haris Skokos) \\ SOME OPEN POINTS IN NONEXTENSIVE STATISTICAL MECHANICS}

\author{Constantino Tsallis
}

\address{Centro Brasileiro de Pesquisas Fisicas \\and National Institute of Science and Technology for Complex Systems\\ 
Rua Xavier Sigaud 150, 22290-180 Rio de Janeiro-RJ, BRAZIL\\
and\\
Institut fur Theoretische Physik III, Justus-Liebig-Universita¨t Giessen, 35392 Giessen, GERMANY\\
%Santa Fe Institute, 1399 Hyde Park Road, Santa Fe, NM 87501, USA\\
tsallis@cbpf.br
}

\maketitle

\begin{history}
\received{(to be inserted by publisher)}
\end{history}

\begin{abstract}
We present and discuss a list of some interesting points that are currently open in nonextensive statistical mechanics. Their analytical, numerical, experimental or observational advancement would naturally be very welcome. 
\end{abstract}


\section{Introduction}
\begin{figure}[h]
\begin{center}
\psfig{file=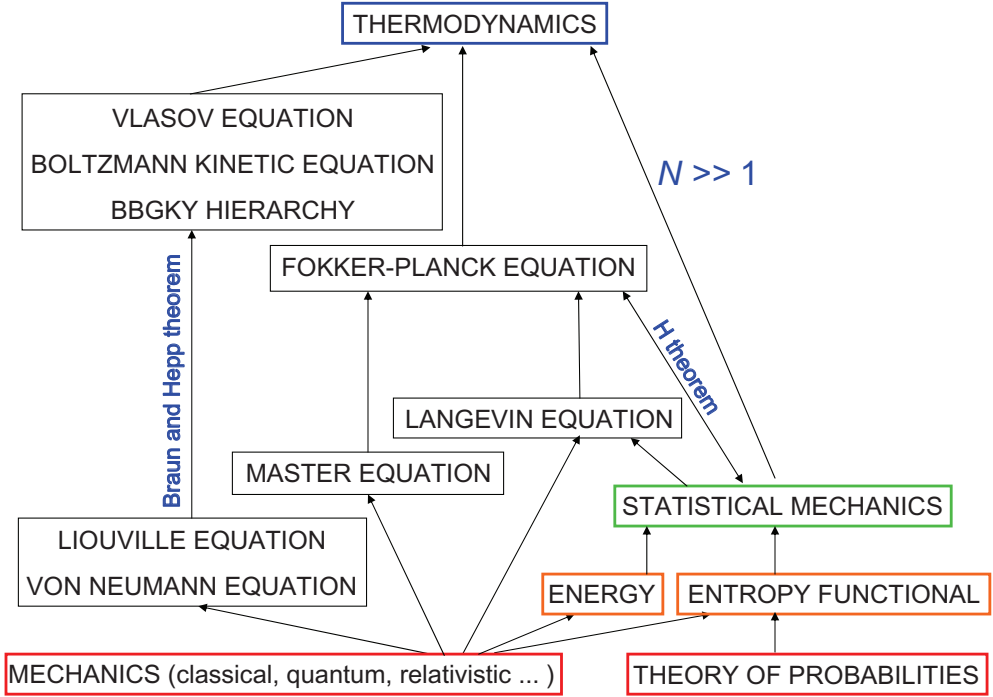,width=4in,angle=-90} %100 percent
\end{center}
\caption{Relevant connections between microscopic (bottom; red boxes) and macroscopic (top; blue box) descriptions through mesoscopic (middle; orange, green and black boxes) ones.}
\label{figconnections}
\end{figure}

\noindent Statistical mechanics is one of the pillars of contemporary physics. Indeed, it provides in principle the link between the {\it microscopic} and the {\it macroscopic} description and understanding of the world through a rich variety of {\it mesoscopic} instances. More precisely, starting from mechanics (classical, quantum, relativistic) and electromagnetism, it incorporates appropriate concepts of the theory of probabilities, and finally leads, for large enough systems, to thermodynamics. It does so through various intermediate descriptions such as those involving master, Langevin, and Fokker-Planck equations. See Fig. \ref{figconnections} for a schematic set of connections.  

The statistical mechanical approach of physical phenomena was first introduced by Maxwell, Boltzmann and Gibbs. A crucial step was done by Boltzmann who proposed a specific functional form for the thermodynamic entropy (that had been introduced a few years earlier by Clausius) in terms of the probabilities of the microscopic configurations. Its expression for say a system describable through a discrete set of probabilities $\{p_i\}$ of $W$ possibilities is given by
\begin{equation}
S_{BG}=-k\sum_{i=1}^Wp_i \ln p_i  \;\;\;\;(\sum_{i=1}^W p_i=1)\,,
\end{equation}
where $BG$ stands for {\it Boltzmann-Gibbs}, and $k$ is a positive constant referred to as the Boltzmann constant (which, together with the velocity of light $c$, the Planck constant $h$ and the gravitational constant $G$, constitutes a minimal set of universal physical constants); for convenience it is sometimes taken $k=1$. The important particular instance where probabilities are all equal (i.e., $p_i=1/W$, $\forall i$) leads to
\begin{equation}
S_{BG}=k\ln W \,,
\end{equation}
carved on Boltzmann's grave stone in the Central Cemetery in Vienna. It is straightforward to verify that in all cases $S_{BG} \ge 0$.

For classical systems, this entropy takes the form
\begin{equation}
S_{BG}=-k\int dx \,p(x) \ln p(x)  \;\;\;\;(\int dx\, p(x)=1)\,,
\end{equation}
where $x$ denotes a generic point in the full phase space of the system (Gibbs' $\Gamma$ space, having typically a dimension equal to $2dN$, where $d$ is the space dimension where the system evolves). This expression cannot be used for too thin distributions, since they would lead to negative entropies. For example, if $p(x)=1/a$ ($a>0$) within a one-dimensional (dimensionless) interval whose width is $a$, and zero otherwise, the entropy will be given by $S_{BG}=k \ln a$. Clearly, this expression is thermodynamically non admissible for $a<1$. This is a well known fact, and is currently considered to reflect the deep quantum nature of the real world.  
 
For quantum systems, the Boltzmann-Gibbs entropy takes the form 
\begin{equation}
S_{BG}=-k \,Tr \rho \ln \rho  \;\;\;\;(Tr \rho=1)\,,
\end{equation} 
where $\rho$ denotes the density matrix of the system. In this form, the BG entropy is frequently referred to as the von Neumann entropy. 
 
The BG entropy and its associated statistical mechanics provide an extremely useful tool for studying a wide variety of physical systems. However, {\it not all} (see, for instance, \cite{Gibbs1902,Fermi1936}). 
Indeed, as suggested since at least 1988 \cite{Tsallis1988}, more general or different entropic functionals become necessary for the statistical mechanics of other, more complex, systems.

The systems for which it is certainly appropriate to apply the BG entropy and consistently associated theories can be loosely\footnote{By {\it loosely} we refer to the fact that, in amazingly many cases, necessary and/or sufficient conditions are not available on rigorous grounds.} characterized by  
short-range space-time correlations, Markovian processes (short memory), additive noise, strong chaos (positive maximal Lyapunov exponent), ergodic dynamics, continuous (Euclidean or Riemannian) geometry for the dynamical occupation of phase space, short-range many-body interactions, weakly quantum-entangled subsystems, linear/homogeneous Fokker-Planck equations, Gausssian distributions. Such systems neatly benefit from the {\it additivity}\footnote{An entropy $S$ is said additive \cite{Penrose1970} if, for two probabilistically independent systems $A$ and $B$ (i.e., for discrete systems, $p_{ij}^{A+B}=p_i^A p_j^B$, $\forall (i,j)$), $S(A+B)=S(A)+S(B)$. This property is straightforwardly satisfied by $S_{BG}$.} of the BG entropy, and typically yield exponential dependences. For example, to start with, the maximization of $S_{BG}$ under appropriate constraints yields, for the canonical ensemble,  the celebrated BG weight itself:
\begin{equation}
p_i^{(BG)} \propto e^{- \beta E_i} \,,
\end{equation} 
where $E_i$ is the energy of state $i$ of a Hamiltonian system satisfying specific boundary conditions, and $\beta = 1/kT$.

There are, however, complex natural, artificial and social systems which, in contrast with the above, can be loosely characterized by 
long-range space-time correlations, non-Markovian processes (long memory), additive and multiplicative noises, weak chaos (vanishing maximal Lyapunov exponent), nonergodic dynamics, hierarchical (typically multifractal) geometry for the dynamical occupation of phase space, long-range many-body interactions, strongly quantum-entangled subsystems, nonlinear/inhomogeneous Fokker-Planck equations, non-Gausssian distributions. A quite wide class among these (though surely {\it not all}) can be handled with the entropy \cite{Tsallis1988}
\begin{equation}
S_{q}=k \frac{1-\sum_{i=1}^W p_i^q}{q-1}  \;\;\;\;(\sum_{i=1}^W p_i=1; \,q \in {\cal R};\, S_1=S_{BG})\,,
\label{qentropy}
\end{equation}
and its analogous continuous and quantum versions. For the particular case of equal probabilities we obtain
\begin{equation}
S_{q}=k \ln_q W\,,
\end{equation}
where
\begin{equation}
\ln_q z \equiv \frac{z^{1-q}-1}{1-q} \;\;\;(z \in {\cal R_+}; \ln_1z=\ln z) \,.
\label{qlogarithm}
\end{equation}
By using this function, entropy (\ref{qentropy}) can be conveniently rewritten as follows:
\begin{equation}
S_{q}=k \sum_{i=1}^W p_i \ln_q \frac{1}{p_i}
\label{qentropy2}
\end{equation}

If $A$ and $B$ are two probabilistically independent systems, we straightforwardly verify that
\begin{equation}
\frac{S_q(A+B)}{k}=\frac{S_q(A)}{k}+\frac{S_q(B)}{k}+(1-q)\frac{S_q(A)}{k}\frac{S_q(B)}{k} \,,
\end{equation} 
which is a direct consequence of the property
\begin{equation}
\ln_q (xy)=\ln_q x + \ln_q y + (1-q) (\ln_q x) (\ln_q y).
\label{pseudoadditive}
\end{equation}
Therefore $S_q$ is {\it nonadditive} for $q \ne 1$ (expressions such as {\it subadditive} and {\it superadditive} are occasionally used to refer to the $q>1$ and $q<1$ cases).

The respective domains of applicability of $S_{BG}$ and of $S_q$ ($q \ne 1$) can be simply characterized through the equal probability case of a large system, i.e., having a number of elements $N>>1$. More precisely, if $W(N) \propto \mu^N\,(\mu >1)$, thermodynamical {\it extensivity} (i.e., $S(N) \propto N$ in the $N \to\infty$ limit) is satisfied by $S_{BG}$. If, in contrast, we have a system constituted by strongly correlated elements such that $W(N) \propto N^\rho \, (\rho>0)$, we verify that $S_q(N) \propto N$ for $q=1-1/\rho<1$. In other words, it is the requirement that the entropy satisfies thermodynamic extensivity which determines the appropriate value of the index $q$ to be used. The main properties of such systems typically exhibit asymptotic power-laws. For example, in what concerns the distribution of energies of a canonical system (i.e., in thermal contact with a thermostat), the extremization of $S_q$ under appropriate constraints \cite{TsallisMendesPlastino1998} (see also \cite{CuradoTsallis1991}) yields
\begin{equation}
p_i^{(q)} \propto e_q^{- \beta E_i} \;\;\;(p_i^{(1)} \propto e^{- \beta E_i}) \,,
\end{equation}       	        
where $\beta$ is related with the temperature $T$, and the $q$-exponential function is defined as the inverse of the $q$-logarithmic function (\ref{qlogarithm}), i.e.,
\begin{equation}
e_q^{z} \equiv [1+(1-q)z]_+^{1/(1-q)} \;\;\;(z \in {\cal R}; e_1^z=e^z) \,,
\end{equation} 
with $[(...)]_+=(...)$ if $(...)>0$, and zero otherwise.

Let us notice that, if we extremize say $S_{2-q}$ instead of $S_q$, we obtain
\begin{equation}
p_i^{(q)} \propto e_{2-q}^{- \beta E_i} \,,
\end{equation} 
which interchanges the $q<1$ and the $q>1$ intervals. We come back onto this point later on. 

The nonadditive entropy $S_q$ and its associated nonextensive statistical mechanics \cite{TsallisBrigatti2004,Tsallis2009a,Tsallis2009b,BibliographyNEXT} have already received a large number of theoretical, experimental, observational and computational verifications, including the prediction of the index (indices) $q$ from first principles\footnote{By {\it first principles} we mean from the set of probabilities of the microscopic configurations and its corresponding dynamics. For example, whenever the system is a mechanical conservative one, from first principles it is meant from the Hamiltonian or from the Lagrangian of its elementary constituents.} for some systems, and its prediction in terms of mesoscopic quantities for some others. Nevertheless, several interesting, delicate and/or elusive points are still open to basic research. It is the purpose of the present paper to review several (inter-related) among them, and to indicate their present status of understanding.  

Before introducing the present list of open points, and in order to clarify what kind of natural, artificial and social phenomena we have in mind, let us mention several of such phenomena that have already been reported in the literature in the frame of the nonadditive entropy $S_q$ and its associated nonextensive statistical mechanics.  

Among others we have (i) The velocity distribution of (cells of) {\it Hydra viridissima} follows
a $q=3/2$ probability distribution function (PDF) \cite{UpadhyayaRieuGlazierSawada2001}; (ii) The velocity distribution of (cells of) {\it
Dictyostelium discoideum} follows a $q=5/3$ PDF in the vegetative
state and a $q=2$ PDF in the starved state \cite{Reynolds2010}; (iii) The velocity
distribution in defect turbulence \cite{DanielsBeckBodenschatz2004}; (iv) The velocity
distribution of cold atoms in a dissipative optical lattice \cite{DouglasBergaminiRenzoni2006}; (v)
The velocity distribution during silo drainage \cite{ArevaloGarcimartinMaza2007a,ArevaloGarcimartinMaza2007b}; (vi) The
velocity distribution in a driven-dissipative 2D dusty plasma, with
$q=1.08\pm0.01$ and $q=1.05\pm 0.01$ at temperatures of $30000 \,K$
and $61000\, K$ respectively \cite{LiuGoree2008}; (vii) The spatial (Monte Carlo)
distributions of a trapped $^{136}Ba^+$ ion cooled by various
classical buffer gases at $300\,K$ \cite{DeVoe2009}; (viii) The distributions of
price returns and stock volumes at the stock exchange, as well as the volatility smile  \cite{Borland2002a,Borland2002b,OsorioBorlandTsallis2004,Queiros2005}; (ix) The
distributions of returns of magnetic field fluctuations in the solar
wind plasma as observed in data from  Voyager 1 \cite{BurlagaVinas2005} and from
Voyager 2 \cite{BurlagaNess2009}; (x) The distributions of returns in the Ehrenfest's
dog-flea model \cite{BakarTirnakli2009,BakarTirnakli2010}; (xi)The distributions of returns  in the
coherent noise model \cite{CelikogluTirnakliQueiros2010}; (xii) The distributions of returns of the
avalanche sizes in the self-organized critical
Olami-Feder-Christensen model, as well as in real earthquakes \cite{CarusoPluchinoLatoraVinciguerraRapisarda2007};
(xiii) The distributions of angles in the $HMF$ model \cite{MoyanoAnteneodo2006}; (xiv)
The distribution of stellar rotational velocities in the Pleiades \cite{CarvalhoSilvaNascimentoMedeiros2008}; (xv) The relaxation in various paradigmatic spin-glass
substances through neutron spin echo experiments \cite{PickupCywinskiPappasFaragoFouquet2009}; (xvi) Various
properties directly related with the time dependence of the width of
the ozone layer around the Earth \cite{FerriReynosoPlastino2010}; (xvii) The distribution of
transverse momenta in high energy collisions of electron-positron,
proton-proton, and heavy nuclei (e.g., Pb-Pb and Au-Au) \cite{BediagaCuradoMiranda2000,WilkWlodarczyk2009,BiroPurcselUrmossy2009,CMS1,CMS2,PHENIX,ShaoYiTangChenLiXu2010}, the flux of solar neutrinos \cite{KaniadakisLavagnoQuarati1996}, and the energy distribution of cosmic rays \cite{TsallisAnjosBorges2003}; (xviii)
Various properties for conservative and dissipative nonlinear dynamical systems \cite{LyraTsallis1998,BorgesTsallisAnanosOliveira2002,AnanosTsallis2004,BaldovinRobledo2004,MayoralRobledo2005,PluchinoRapisardaTsallis2007,PluchinoRapisardaTsallis2008,MiritelloPluchinoRapisarda2009,LeoLeoTempesta2010}; (xix) The degree distribution of (asymptotically) scale-free
networks \cite{WhiteKejzarTsallisFarmerWhite2006,ThurnerKyriakopoulosTsallis2007}; (xx) Tissue radiation response \cite{Sotolongo-GrauRodriguez-PerezAntoranzSotolongo-Costa2010}; (xxi) Overdamped motion of interacting particles \cite{AndradeSilvaMoreiraNobreCurado2010}\footnote{Many other phenomena have been looked at along similar lines (e.g., biological evolution \cite{TamaritCannasTsallis1998}, turbulence in electron plasma \cite{AnteneodoTsallis1997}).}.

\section{Some Open Points}

\subsection{$q$-generalized Lyapunov spectrum and Pesin identity}

\subsubsection{$q$-generalized Lyapunov spectrum}

Let us illustrate this point through a simple example, namely a one-dimensional dissipative map, e.g., the z-logistic family
\begin{equation}
x_{t+1}=1-a|x_t|^z \;\;\;(t=0,1,2,3,...; \,-1 \le x_t \le 1; \,z>1; \,0 \le a \le 2)
\label{zlogistic}
\end{equation}
The particular case $z=2$ corresponds to the standard logistic map; the $z \to 1$ limit corresponds to the tent map. These maps metrically, but not topologically, differ for different values of $z$. For the simple $z=2$ case, and with $y \equiv x+1/2$, we obtain the traditional form
\begin{equation}
y_{t+1}=\mu \, y_t(1-y_t)     \;\;\;\;( 0 \le \mu \le 4; 0 \le y_t \le 1) \,.     
\end{equation}

For $a$ increasing above zero, a succesion of fixed points and fixed cycles occur, separated by doubling-period bifurcations. These bifurcations accumulate as $a$ approaches a special point, $a_c(z)$, the first {\it edge of chaos}. For $z=2$ it is $a_c(2)=1.40115518909...$   

The sensitivity to the initial conditions $\xi$ for a one-dimensional dynamical system is defined as follows:
\begin{equation}
\xi(t) \equiv \lim_{\Delta x(0) \to 0} \frac{\Delta x(t)}{\Delta x(0)} \;\;\;(\xi(0)=1)\,, 
\end{equation}
where $x$ denotes the phase space variable. The system is said {\it strongly chaotic} (or simply {\it chaotic}) if $\xi$ {\it exponentially} diverges with time. In such cases we can define the Lyapunov exponent $\lambda$ through
\begin{equation}
\xi(t) \sim e^{\lambda\, t} \,,
\end{equation}
or, more precisely, through
\begin{equation}
\lambda \equiv \lim_{t\to\infty} \frac{\ln \xi(t)}{t} \,.
\end{equation}

At the edge of chaos, $\lambda$ vanishes and $\xi$ increases slowly with $t$, in fact algebraically at large enough values of $t$. For $a$ increasing above $a_c(z)$, $\lambda$ greatly oscillates in a complex manner, being however positive for most of the values of $a$. 
The sensitivity $\xi$ is, in fact, quite generically expected to satisfy
\begin{equation}
\frac{d\xi}{dt}=\lambda_{q_{sen}}\xi^{\,q_{sen}} \;\;\;(\xi(0)=1)\,,
\label{qexponentialequation}
\end{equation}
hence \cite{TsallisPlastinoZheng1997,BaldovinRobledo2002a,BaldovinRobledo2002b,Robledo2006} 
\begin{equation}
\xi(t) = e_{q_{sen}}^{\lambda_{q_{sen}} \,t} =[1+(1-q_{sen})\lambda_{q_{sen}}\,t]^{\frac{1}{1-q_{sen}}} \,,\label{qexponentialxi}
\end{equation}
where $q_{sen}=1$ if the Lyapunov exponent $\lambda_1 \equiv \lambda \ne 0$ ({\it strongly sensitive} if $\lambda_1>0$, and {\it strongly insensitive} if $\lambda_1<0$), and $q_{sen} \ne 1$ otherwise; {\it sen} stands for {\it sensitivity}.  At the edge of chaos, $q_{sen} < 1$ ({\it weakly sensitive}), and at both the period-doubling and tangent bifurcations, $q_{sen} > 1$ ({\it weakly insensitive}). The case $q_{sen}<1$  yields, in (\ref{qexponentialxi}), a power-law behavior $\xi \propto t^{1/(1-q_{sen})}$ in the limit $t \to\infty$. This power-law asymptotics were since long known in the literature   \cite{GrassbergerScheunert1981,SchneiderPolitiWurtz1987,AnaniaPoliti1988,HataHoritaMori1989,MoriHataHoritaKobayashi1989}. The case $q_{sen}<1$ is in fact more complex than indicated in Eqs. (\ref{qexponentialequation}) and (\ref{qexponentialxi}). These equations only reflect the {\it maximal} values of an entire family, fully (and not only asymptotically) described in \cite{Robledo2006,MayoralRobledo2005}. 
%See Figs. \ref{BaldovinRobledo2002b} and \ref{BaldovinRobledo2004} from \cite{BaldovinRobledo2002b}.

{\it The rigorous necessary and sufficient conditions for behaviors such as those indicated in Eqs. (\ref{qexponentialequation}) and (\ref{qexponentialxi}), for generic conservative and dissipative nonlinear maps, stand at present as an open problem.} 

\subsubsection{Pesin-like identity}

Let us now focus on the time evolution of the entropy of the above illustrative maps. We shall illustrate with the map defined in Eq. (\ref{zlogistic}). We divide the phase space $x \in [-1,1]$ in $W$ equally spaced little intervals denoted with $i=1,2,3,...,W$(with $W>>1$), choose one of them, and put within $M$ (randomly chosen) initial conditions (with $M>>1$; typically $M \simeq 10 \times W$). At time $t$ we have in the $i^{th}$ interval $M_i(t)$ points; naturally $\sum_{i=1}^W M_i(t)=M$. We then define a set of probabilities through $p_i(t) \equiv M_i(t)/M$ (hence $\sum_{i=1}^W p_i(t)=1$), which enables the calculation of the entropy (we take $k=1$)
\begin{equation}
S_q(t)= \frac{1-\sum_{i=1}^W [p_i(t)]^q}{q-1}\,.
\end{equation}
We next define the {\it entropy production} (per unit time) as follows:
\begin{equation}
K_q \equiv \lim_{t \to\infty} \lim_{W \to\infty} \lim_{M\to\infty} \frac{S_q(t)}{t} \,.
\end{equation}
There typically exists an unique value of $q$, noted $q_{production}$, such that $K_{q_{production}}$ is {\it finite}. For $q> q_{production}$ ($q< q_{production}$), $K_q$ vanishes (diverges). For various one-dimensional dissipative maps (and also two-dimensional conservative ones) we verify that
\begin{equation}
q_{production}=q_{sen}
\label{qequality}
\end{equation}
and
\begin{equation}
K_{q_{production}}=\lambda_{q_{sen}}\,.
\label{qpesin}
\end{equation}
For strongly chaotic maps, i.e. with $\lambda_1>0$, we verify that
\begin{equation}
q_{production}=q_{sen}=1 \,,
\end{equation}
and
\begin{equation}
K_1=\lambda_1 >0.
\end{equation}   
This last equality can be seen as a Pesin-like one.

\begin{figure}[h]
\begin{center}
\hspace{0.3cm}
\psfig{file=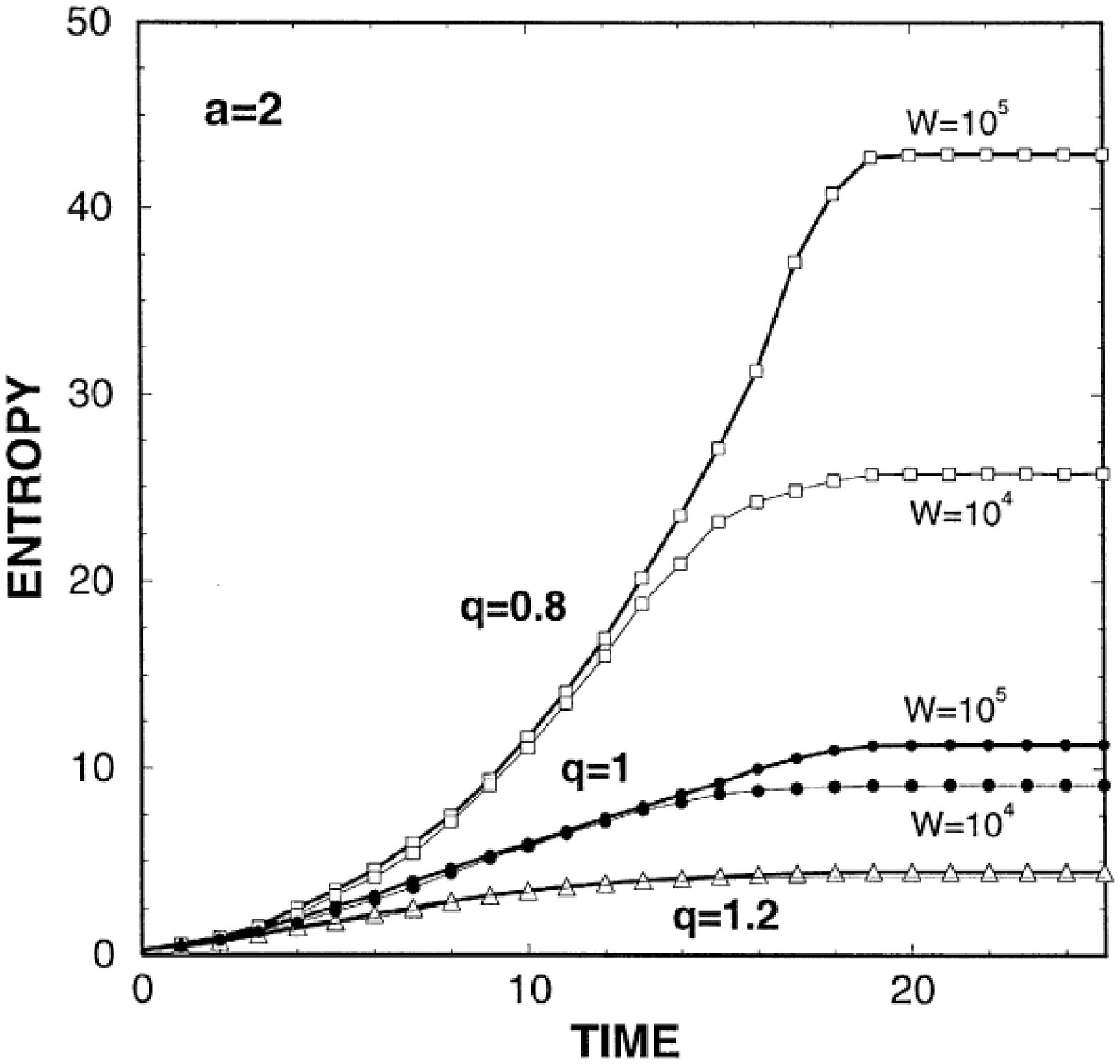,width=4.2in,angle=-90}
\psfig{file=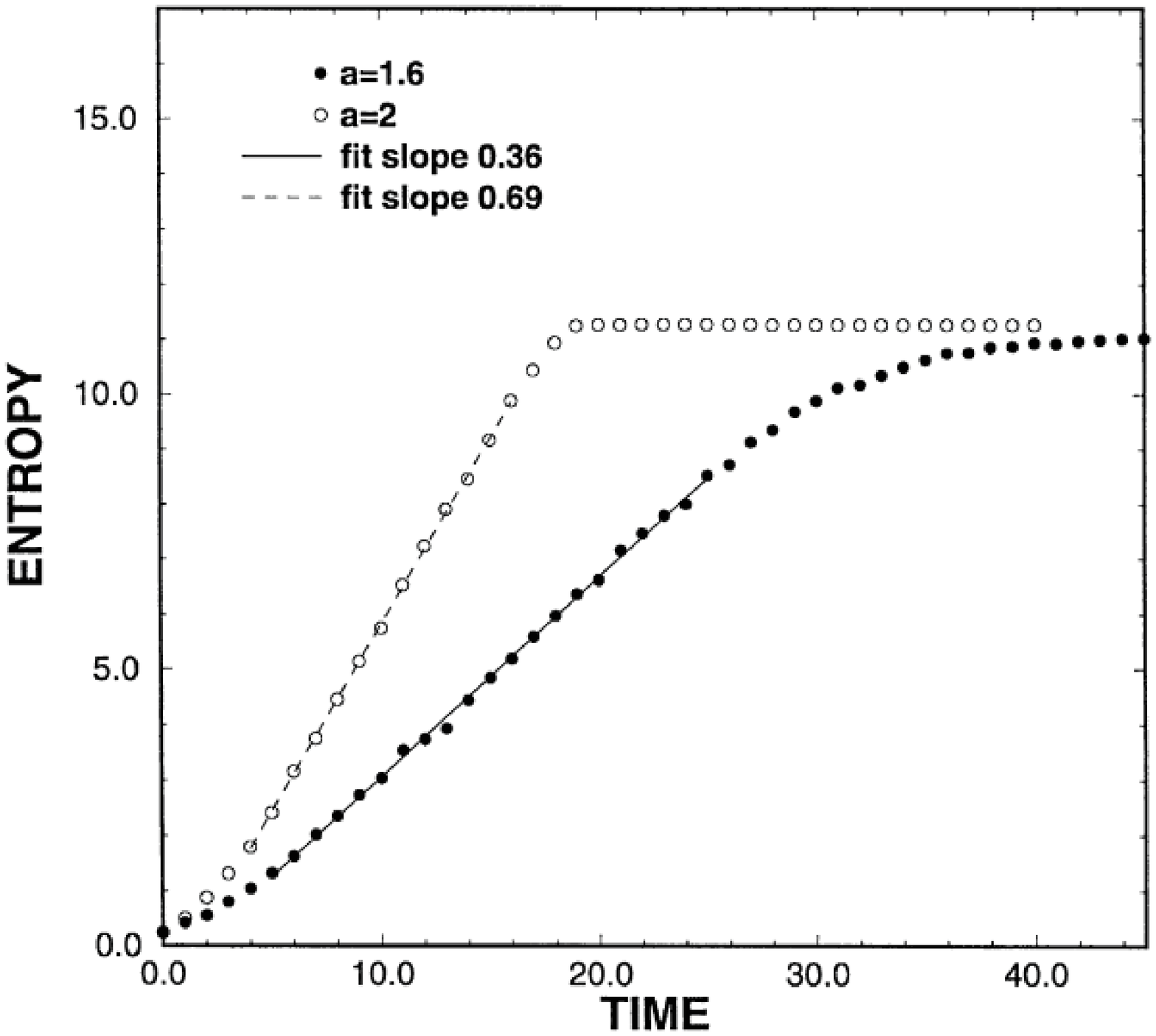,width=3.6in,angle=-90}
\end{center}
\caption{Time evolution of the BG entropy for typical values of $a$ of the $z=2$ logistic map, corresponding to positive Lyapunov exponents. {\it Top}: We notice that only for $q=1$ we observe a {\it linear} intermediate behavior of the entropy before saturation. This linear region is larger for larger $W$. {\it Bottom}: The entropy production $K_{BG}$ per unit time decreases when the value of $a$ corresponds to a smaller value of the Lyapunov exponent. From \cite{LatoraBarangerRapisardaTsallis2000}.}
\label{mapchaos}
\end{figure}

For weakly chaotic ones (typically at the edge of chaos), i.e., with $\lambda_1=0$, we verify that
\begin{equation}
q_{production}=q_{sen}<1 \,,
\end{equation}
and
\begin{equation}
K_{q_{production}}=\lambda_{q_{sen}} >0.
\end{equation}   
This last equality can be seen as the $q$-generalization of the Pesin-like equality. See Fig. \ref{mapchaos}

{\it The rigorous necessary and sufficient conditions for equalities such as those indicated in Eqs. (\ref{qequality}) and (\ref{qpesin}), for generic conservative and dissipative nonlinear maps, stand at present as an open problem.} 

\subsection{Geometry of occupation of phase space, Hilbert space}
                     
\subsubsection{Additivity versus extensivity of the entropy}

The {\it additivity} of an entropy only depends on the mathematical functional which expresses the macroscopic (thermodynamical) entropy in terms of its basic probabilities. Therefore, as already said, the BG entropy $S_{BG}$ is additive. So is the so-called Renyi entropy $S_q^R$ (useful in the geometrical description of hierarchical structures such as multifractals), defined (for the discrete case) as follows
\begin{equation}
S_q^R = \frac{\ln \Bigl( \sum_{i=1}^W p_i^q \Bigr)}{1-q} \;\;\;(q \in {\cal R}; S_1^R=S_{BG}) \,.
\end{equation}
Indeed, it can be straightforwardly proved that, if $A$ and $B$ are two probabilistically independent systems, then $S_q^R(A+B)=S_q^R(A)+S_q^R(B)$. 

In contrast, the {\it extensivity} of an entropy is a more subtle concept, and it depends not only on its functional form but also on the specific system under consideration, i.e., on its microscopic probabilistic correlations. Consequently, while the expression ``the BG entropy is additive" is definitively correct, the expression  ``the BG entropy is extensive" is incorrect. The correct expression would be ``the BG entropy of this class of systems is extensive". 

Consider a system $\Sigma \equiv A_1+A_2+...+A_N$ constituted by $N$ (not necessarily independent) identical elements or subsystems $\{A_j\}$. An entropy $S$ of that system is {\it extensive} if $0<\lim_{N \to\infty} \frac{S(N)}{N}< \infty$, i.e., if
\begin{equation}
S(N) \propto N \;\;\;(N \to\infty) \,.
\end{equation}

The important difference between additivity and extensivity can be illustrated through probabilistic systems of $N$ identical binary random variables. If the variables are {\it distinguishable}, we may represent them in the following triangular form:  

\begin{table}[htbp]
\begin{flushleft}
~~~~~~~~~~~~~~$(N=0)$        ~~~~~~~~~~~~~~~~~~~~~~         ~~~~$1 \times1$~~~~\\
~~~~~~~~~~~~~~$(N=1)$        ~~~~~~~~~~~~~~~~~~~~~~$1 \times r_{1,0}$~~$1 \times r_{1,1}$~~~\\ 
~~~~~~~~~~~~~~$(N=2)$        ~~~~~~~~~~~~~~~~~$1\times r_{2,0}$~~$2 \times r_{2,1}$~~$1\times r_{2,2}$~~\\ 
~~~~~~~~~~~~~~$(N=3)$        ~~~~~~~~~~~$1\times r_{3,0}$~~$3\times r_{3,1}$~~$3\times r_{3,2}$~~$1\times r_{3,3}$~\\ 
~~~~~~~~~~~~~~$(N=4)$    ~~~~~      $1\times r_{4,0}$~~$4\times r_{4,1}$~~$6\times r_{4,2}$~~$4\times r_{4,3}$~~$1\times r_{4,4}$\\
\end{flushleft}
%\vspace{-0.5cm}
\caption{Merging of the Pascal triangle ....  
}
\label{general}
\end{table}
where $0 \le r_{N,n} \le 1$ and 
\begin{equation}
\sum_{n=0}^N \frac{N!}{(N-n)!\,n!}\,r_{N,n}=1 \;\;\;(\forall N)
\end{equation}

If the distinguishable variables are {\it independent} we obtain the particular case
\begin{equation}
r_{N,n}=p^{N-n}(1-p)^n \;\;\;(0 \le p \le 1)
\end{equation}

\begin{table}[htbp]
\begin{flushleft}
~~~~~~~~~~~~~~$(N=0)$        ~~~~~~~~~~~~~~~~~~~~~~         ~~~~$1 \times1$~~~~\\
~~~~~~~~~~~~~~$(N=1)$        ~~~~~~~~~~~~~~~~~~~~~~$1 \times p$~~$1 \times(1-p)$~~~\\ 
~~~~~~~~~~~~~~$(N=2)$        ~~~~~~~~~~~~~~~~~$1\times p^2$~~$2 \times p(1-p)$~~$1\times (1-p)^2$~~\\ 
~~~~~~~~~~~~~~$(N=3)$        ~~~~~~~~~~~$1\times p^3$~~$3\times p^2(1-p)$~~$3\times p(1-p)^2$~~$1\times(1-p)^3$~\\ 
~~~~~~~~~~~~~~$(N=4)$    ~~~~~      $1\times p^4$~~$4\times p^3(1-p)$~~$6\times p^2(1-p)^2$~~$4\times p(1-p)^3$~~$1\times (1-p)^4$\\
\end{flushleft}
\caption{Merging of the Pascal triangle......  }
\label{independent}
\end{table}

If $p=1/2$, this triangle becomes 

\begin{table}[htbp]
\begin{flushleft}
~~~~~~~~~~~~~~$(N=0)$        ~~~~~~~~~~~~~~~~~~~~~~         ~~~~$1 \times1$~~~~\\
~~~~~~~~~~~~~~$(N=1)$        ~~~~~~~~~~~~~~~~~~~~~~$1 \times \frac{1}{2}$~~$1 \times\frac{1}{2}$~~~\\ 
~~~~~~~~~~~~~~$(N=2)$        ~~~~~~~~~~~~~~~~~$1\times\frac{1}{4}$~~$2 \times \frac{1}{4}$~~$1\times \frac{1}{4}$~~\\ 
~~~~~~~~~~~~~~$(N=3)$        ~~~~~~~~~~~$1\times\frac{1}{8}$~~$3\times\frac{1}{8}$~~$3\times\frac{1}{8}$~~$1\times\frac{1}{8}$~\\ 
~~~~~~~~~~~~~~$(N=4)$    ~~~~~      $1\times\frac{1}{16}$~~$4\times\frac{1}{16}$~~$6\times\frac{1}{16}$~~$4\times\frac{1}{16}$~~$1\times\frac{1}{16}$\\
\end{flushleft}
%\vspace{-0.5cm}
\caption{Merging of the Pascal triangle ....  
}
\label{independentequal}
\end{table}

In the presence of correlations, the generic triangle can take various forms characterized by the sets $\{r_{N,n} \} \;, \forall N$. For example we have the Leibnitz triangle. 
\begin{table}[htbp]
\begin{flushleft}
~~~~~~~~~~~~~~$(N=0)$        ~~~~~~~~~~~~~~~~~~~~~~         ~~~~$1 \times1$~~~~\\
~~~~~~~~~~~~~~$(N=1)$        ~~~~~~~~~~~~~~~~~~~~~~$1 \times \frac{1}{2}$~~$1 \times\frac{1}{2}$~~~\\ 
~~~~~~~~~~~~~~$(N=2)$        ~~~~~~~~~~~~~~~~~$1\times\frac{1}{3}$~~$2 \times \frac{1}{6}$~~$1\times \frac{1}{3}$~~\\ 
~~~~~~~~~~~~~~$(N=3)$        ~~~~~~~~~~~$1\times\frac{1}{4}$~~$3\times\frac{1}{12}$~~$3\times\frac{1}{12}$~~$1\times\frac{1}{4}$~\\ 
~~~~~~~~~~~~~~$(N=4)$    ~~~~~      $1\times\frac{1}{5}$~~$4\times\frac{1}{20}$~~$6\times\frac{1}{30}$~~$4\times\frac{1}{20}$~~$1\times\frac{1}{5}$\\
\end{flushleft}
\caption{Merging of the Pascal triangle ....  
}
\label{leibnitz}
\end{table}
Its generic term is given by
\begin{equation}
r_{N,n}=\frac{(N-n)!\,n!}{N!}\frac{1}{N+1} \,.
\end{equation}

The Leibnitz triangle satisfies the following remarkable property (from now on referred to as the {\it triangle Leibnitz rule}):
\begin{equation}
r_{N,n}+r_{N,n=1}=r_{N-1,n} \,.
\end{equation}
This property implies probabilistic {\it scale-invariance}. Indeed, it implies that the marginal probabilities of a $N$-system coincide with the joint probabilities of a $(N-1)$-system. 
 
Let us now turn back to the entropy. For a wide class of systems (which includes all the above probabilistic triangles) a value of $q$ exists, noted $q_{entropy}$, such that $S_{q_{entropy}}$ is extensive. For all the above examples it is $q_{entropy}=1$, i.e., $S_{BG}(N) \propto N \;\;(N \to\infty)$.

Let us now severely restrict the admissible probabilistic region (defined as the set of configurations whose probability is strictly positive). We are referring to cases where the total number, noted $W_{eff}(N)$ ({\it eff} stands for {\it effective}), of admissible configurations is much smaller than $W(N)$ for $N>>1$. In other words, cases where $\lim_{N \to\infty} W_{eff}(N)/W(N)=0$. Two such examples are indicated in what follows (see details in \cite{TsallisGellMannSato2005}). These triangles have nonzero probabilities only along a (left) strip whose width is denoted $d+1$. The two examples here respectively correspond to $d=1$ and $d=2$.

\begin{table}[htbp]
~~~~~~~~~~~~~~~~~~~~$(N=0)$~~~~~~~~~~~~~~~~~~~~~~~~~~$1 \times 1$~~~~~~~~~~~~~~~~~~~~~~~~~~~~~~~

~~~~~~~~~~~~~~~~~~~~$(N=1)$~~~~~~~~~~~~~~~~~~~$1 \times \frac{1}{2}$~~$1 \times \frac{1}{2}$~~~~~~~~~~~~~~~~~~~~~~~~  

~~~~~~~~~~~~~~~~~~~~$(N=2)$~~~~~~~~~~~~~~$1 \times \frac{1}{2}$~~$2 \times \frac{1}{4}$~~$1 \times 0$~~~~~~~~~~~~~~~~~~~~    
 
~~~~~~~~~~~~~~~~~~~~$(N=3)$~~~~~~~~~$1 \times \frac{1}{2}$~~$3 \times \frac{1}{6}$~~~$3 \times 0$~~$1 \times 0$~~~~~~~~~~~~~~~
 
~~~~~~~~~~~~~~~~~~~~$(N=4)$~~~$1 \times \frac{1}{2}$~~$4 \times \frac{1}{8}$~~~~$6 \times 0$~~$4 \times 0$~~$1 \times 0$~~~~~~~~~~~~

\vspace{0.5cm}

~~~~~~~~~~~~~~~~~~~~$(N=0)$~~~~~~~~~~~~~~~~~~~~~~~~~~$1 \times 1$~~~~~~~~~~~~~~~~~~~~~~~~~~~~~~~~

~~~~~~~~~~~~~~~~~~~~$(N=1)$~~~~~~~~~~~~~~~~~~$1 \times \frac{1}{2}$~~$1 \times \frac{1}{2}$~~~~~~~~~~~~~~~~~~~~~~~~~  
 
~~~~~~~~~~~~~~~~~~~~$(N=2)$~~~~~~~~~~~~$1 \times \frac{1}{3}$~~$2 \times \frac{1}{6}$~~$1 \times \frac{1}{3}$~~~~~~~~~~~~~~~~~~~    
 
~~~~~~~~~~~~~~~~~~~~$(N=3)$               ~~~~~$1 \times \frac{3}{8}$~$3 \times \frac{5}{48}$~~$3 \times \frac{5}{48}$~~$1 \times 0$~~~~~~~~~~~~~~
 
~~~~~~~~~~~~~~~~~~~~$(N=4)$~                    $1 \times \frac{2}{5}$  $4 \times \frac{3}{40}$  $6 \times \frac{3}{60}$~~~~$4 \times 0$~~$1 \times 0$ ~~~~~~~~~\\
\caption{Anomalous probability sets: $d=1$ ({\it top}), and $d=2$ ({\it bottom}). The left number within parentheses indicates the multiplicity (i.e., Pascal triangle). The right number indicates the corresponding probability. The probabilities, noted $r_{N,n}$,  asymptotically satisfy the Leibnitz rule, i.e., $  \lim_{N\to \infty}\frac{r_{N,n}+r_{N,n+1}}{r_{N-1,n}}=1$ ($\forall n$). In other words, the system is, in this sense, {\it asymptotically scale-invariant}. Notice that the number of triangle elements with nonzero probabilities grows like $N$, whereas that of zero probability grows like $N^2$.
} 
\label{tablePNAS2}
\end{table}

It can be shown that, for this family of triangles, we have $W_{eff}(N)<<W(N)=2^N \;\;\;(N>>1)$, and consistently
\begin{equation}
q_{entropy}= 1 - \frac{1}{d} \;\;\;(d=1,2,3,...)\,.
\end{equation}

\subsubsection{Some many-body physical examples enabling first-principle calculations of $q$}

Quantum entanglement is caused by the intrinsic nonlocality of quantum mechanics. This nonlocality makes the elements of a $N$-body ($N>>1$) system to be strongly correlated, which diminishes considerably the size of the admissible space of microscopic configurations. Depending on the specific system, $S_{BG}(N)$ might be extensive. We consider here what is currently referred to as the {\it block entropy}, i.e., the entropy of a subsystem which is very large and nevertheless much smaller that the entire $N$-system. The elements of this subsystem are quantum-entangled with {\it all} the $N$ elements, but our detector only detects (for whatever reason) the elements of the subsystem. For the subsystem, it frequently happens that its $S_{BG}$ is nonextensive, and therefore {\it inadequate for thermodynamical purposes}. We review here two magnetic examples for which a nontrivial value $q_{entropy}$ exists such that the block entropy $S_{q_{entropy}}$ is extensive, thus reconciling the subsystem entropy with thermodynamics. 

Both examples concern chains of many-body spin systems with short-range interactions at $T=0$, i.e., at their fundamental state. The total $N$-system is in a {\it pure state}, and therefore its entropy vanishes. A block of successive $L$ spins (with $1<<L<<N$) is, however, in a {\it mixed state} and therefore its entropy $S_q(L)$ is different from zero. By  $S_q(L)$ we mean precisely
\begin{equation}
S_q(L)=\frac{1-Tr \rho_L^q}{q-1}\,,
\end{equation}
where
\begin{equation}
\rho_L = Tr_{N-L}\rho_N \,,
\end{equation}
$\rho_N$ being the density matrix of the $N$-system. We have that $Tr \rho_N^2=1$ (i.e., a pure state, hence $S_q(N)=0$), whereas $Tr \rho_L^2<1$ (i.e., a mixed state, hence $S_q(L)>0$).

The first example is a pure ferromagnet with spin 1/2 anisotropic first-neighbor interactions in the $XY$ plane in the presence of a transverse magnetic field at its critical value (a $T=0$ quantum phase transition from the paramagnetic to the ferromagnetic state). It contains the Ising  ($XY$ isotropic) ferromagnet as the {\it central charge} $c=1/2$ ($c=1$) particular case. We analytically find \cite{CarusoTsallis2008} that
\begin{equation}
q_{entropy}= \frac{\sqrt{9+c^2}-3}{c} \;\;\;(c \ge 0)\,.
\end{equation}

The second example is a random magnet of the Heisenberg type with spins $S$, in the absence of any external field. It is numerically found \cite{SaguiaSarandy2010} that
\begin{equation}
q_{entropy}= 1-\frac{1.67}{c}=1-\frac{1.67}{\ln (2S+1)} \;\;\;(c \ge 0)\,.
\end{equation}
 
\begin{figure}[h]
\begin{center}
\hspace{-0.5cm}
\psfig{file=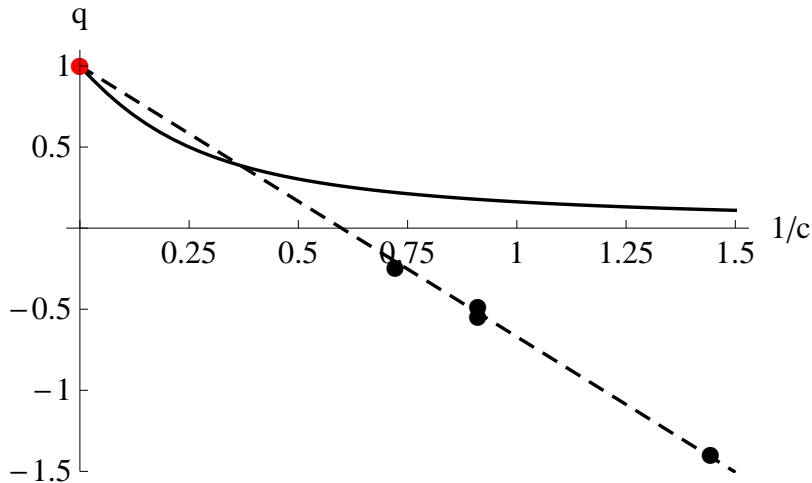,width=4.2in,angle=0}
\end{center}
\caption{Curves of $q$ versus $1/c$. The red dot corresponds to the BG limit. The continuous (dashed) curve corresponds to the $T=0$ linear chain for the pure (random) magnet in \cite{CarusoTsallis2008} (\cite{SaguiaSarandy2010}).}
\label{entanglement}
\end{figure}

Both expressions for $q_{entropy}$ are represented in Fig. \ref{entanglement}.

Since the block is not in a pure but in a mixed state, it necessarily has a nontrivial energy distribution. Which distribution? It could be given, in the $N \to\infty$ limit, by say
\begin{equation}
\rho_L \propto e_{q_{energy}}^{-\beta_{q_{energy}} \,H_L} \;\;\;(L>>1)\,,
\label{blockdensity}
\end{equation}
where $q_{energy}$ could well be a (relatively simple) function of $q_{entropy}$: 
\begin{equation}
q_{energy}=h(q_{entropy}) \,.
\end{equation}
In the $c \to\infty$ limit one naturally expects the BG limit $q_{energy}=q_{entropy}=1$ to hold (i.e., $h(1)=1$). 

{\it The block density matrix $\rho_L$ stands at present as an open problem. If it turns out to be of the form (\ref{blockdensity}), what would be the values of $q_{energy}$ and of $\beta_{q_{energy}}$? }

\subsection{q-generalized central limit theorems}

\subsubsection{$q$-product}

For reasons that will soon become clear let us introduce a $q$-generalization of the product, {\it the $q$-product} \cite{NivanenLeMehauteWang2003,Borges2004}:
\begin{equation}
x \otimes_q y= [x^{1-q}+y^{1-q}]^{\frac{1}{1-q}} \;\;\;\;(x \otimes_1 y=xy) \,.
\label{qproduct}
\end{equation}
The main reason for defining such a product is that it satisfies the following remarkable property:
\begin{equation}
\ln_q (x \otimes_q y)= \ln_q x + \ln_q y   \,.
\label{qextensivity}
\end{equation}
Similarly to Eq. (\ref{pseudoadditive}), which essentially reflects the nonadditivity of the entropy $S_q$, Eq. (\ref{qextensivity}) reflects its possible extensivity in the presence of strong correlations characterized by $q$. Let us illustrate this fact. Assume that we have a system whose subsystem $A$ ($B$) has $N_A>>1$ ($N_B>>1$) strongly correlated elements such that its nonzero-probability configurations are equally probable, and equal to $1/W(N_A)$ ($1/W(N_B)$). Assume also that the correlations are such that $W(N_A)\propto (N_A)^\rho$ ($W(N_B)\propto (N_B)^\rho$) with $\rho \in {\cal R}$. Then we straightforwardly verify entropic extensivity for $q=1-1/\rho$, i.e., that $S_{1-1/\rho}(N_A + N_B) \propto N_A+N_B$.  

We shall focus here on the $q$-product for the case $q \ge 1$, and $x$ and $y$ non-negative real numbers (see \cite{TsallisQueiros2007} for further details).
It is straightforward to check that this product is commutative, associative, it has an inverse, a unit, and a zero. But there are preliminary indications that there is no associative generalized sum with regard to which the $q$-product is distributive. The argument goes essentially as follows. We first develop the $q$-product as follows (see Appendix A in \cite{Tsallis2009a}):
\begin{equation}
x \otimes_q y=  x y \Bigl\{ 1+(q-1) (\ln x) (\ln y)         
    + \frac{1}{2} (q-1)^2 \, [(\ln^2 x)( \ln y)
+(\ln x) (\ln^2 y)+ (\ln^2 x) (\ln^2 y)]+...\Bigr\}  \,.
\end{equation}
We next assume the existence of a generalized sum $x \,\bar\oplus_q \,y$ whose development would be as follows \footnote{We use $x\, \bar\oplus_q \,y$ instead of simply $x \oplus_q y$ because, for different purposes, the latter is already defined in the literature. It is called $q$-sum and is defined as $x \oplus_q y=x+y+(1-q)xy$. Let us say right away that $x \otimes_q y$ is {\it not} distributive with regard to $x \oplus_q y$, as can be easily verified.}:
\begin{equation}
x \,\bar\oplus_q \,y=  (x+ y) \Bigl\{ 1+(q-1) a_1(x,y)         
    + \frac{1}{2} (q-1)^2 a_2(x,y)+...\Bigr\}  \,,
\end{equation}
where $a_1(x,y)=a_1(y,x)$ and $a_2(x,y)=a_2(y,x)$ are functions to be found by simultaneously imposing (i) distributivity of the $q$-product with regard to $x \,\bar\oplus_q \,y$, and (ii) associativity of this generalized sum. In other words we impose
\begin{equation}
x \otimes_q (y \,\bar\oplus_q \,z)=(x \otimes_q y) \,\bar\oplus_q \, (x \otimes_q z) \,,
\label{distributive}
\end{equation}
and
\begin{equation}
x \,\bar\oplus_q \, (y \,\bar\oplus_q \,z)= (x \,\bar\oplus_q \, y) \,\bar\oplus_q \,z\,.
\label{associative}
\end{equation}
Also, we impose $x \,\bar\oplus_q \,0=0 $, $\forall q$, which immediately implies $a_1(x,0)=a_2(x,0)=0$. By developing in powers of $(q-1)$ both relations (\ref{distributive}) and (\ref{associative}), we obtain strong constraints on $a_1(x,y)$ and $a_2(x,y)$. Although the full discussion remains to be done, a preliminary analysis \cite{CuradoRoditiTsallis2008} suggests that no  such $a_1(x,y)$ and $a_2(x,y)$ can exist. If indeed they do not exist, no algebra exists whose product would be the $q$-product. That would be some kind of new mathematical structure, satisfying all the axioms of an algebra but distributivity. 

{\it The distributivity of the $q$-product stands at present as an open problem. Depending on whether it exists or not, and on what would be the analytical definition of $x \,\bar\oplus_q \,y$, is it possible to construct some structure having some similarity with a vector space?}

\subsubsection{q-Fourier transform and discussion of its inverse}

By using the definition (\ref{qproduct}) we can define the following $q$-generalized Fourier transform ($q$-FT) \cite{UmarovTsallisSteinberg2008,UmarovTsallisGellMannSteinberg2010}:
\begin{equation}
F_q[f(x)](\xi)=\int dx \,e_q^{ix\xi} \otimes f(x) \;\;\;(q \ge 1)
\label{qfourierdefinition}
\end{equation}
For non-negative real $f(x)$ we have that
\begin{equation}
F_q[f(x)](\xi)=\int dx \,e_q^{ix\xi \, [f(x)]^{q-1}}  f(x) \;\;\;(q \ge 1)
\end{equation}
It is clear that $1$-FT recovers the standard Fourier transform (FT), which is a linear integral transform.
It is also clear that, for $q\ne 1$, the $q$-FT is a {\it nonlinear} integral transform \footnote{If $x$ is a variable which carries physical dimensions, it is convenient to define definition (\ref{qfourierdefinition}) as follows: $F_q[f(x)](\xi)=\int d(xf_0) \,e_q^{ix\xi} \otimes (f(x)/f_0)=\int d(xf_0) \,e_q^{ix\xi \, [f(x)/f_0]^{q-1}}  (f(x)/f_0)=\int dx \,e_q^{ix\xi \, [f(x)/f_0]^{q-1}}  f(x) \;\;\;(q \ge 1)$, where $f_0>0$ is a reference value. If $f(0)$ is finite, a simple choice would be to just adopt $f_0=f(0)$. With this generalized definition it follows that $x\xi$ is a pure number, which physically is very convenient. If $x$ is already a pure number, we can of course adopt $f_0=1$.}. 
It has been so defined in order to be closed with regard to the family of the $q$-Gaussian distributions. Let us be explicit. We consider the following $q$-Gaussian \footnote{$q$-Gaussians are referred to with various names in the literature. In the form $f(x) \propto 1/[a^2 + x^2]^\kappa$, they are occasionally called  ``generalized Lorentzians" or ``kappa distributions" in areas such as plasma physics, and ``Barenblatt form" in the area of porous media. Also they recover the Student' s $t$-distributions and the $r$-distributions for special values of $q$. Because of that, in finance, they are loosely referred to as ``Student' s $t$-distributions" even {\it outside} these special values of $q$.}
\begin{equation}
f(x)=G_q(x) \equiv \frac{[1-(1-q)\beta\,x^2]^{\frac{1}{1-q}}}{\int dy \, [1-(1-q)\beta\,y^2]^{\frac{1}{1-q}}}=\frac{e_q^{-\beta x^2}}{\int dy\, e_q^{- \beta y^2}}\;\;\;(\beta>0; \, q<3)\,,
\label{qGaussian1}
\end{equation}
where we remind that the support is infinite for $1 \le q < 3$, and is finite for $q<1$ \footnote{Let us remind that the $q$-Gaussian form is normalizable only for $q<3$. Its variance is finite for $q <5/3$, and diverges for $5/3 \le q <3$. Its $q$-variance, however, remains finite for any $q<3$, i.e., as long as it is normalizable.}. Eq. (\ref{qGaussian1}) can be written explicitly as follows: 
\begin{equation}
G_q(x) = N_q \sqrt{\beta} \, [1-(1-q)\beta\,x^2]^{\frac{1}{1-q}}\;\;\;(\beta>0; \, q<3)\,,
\end{equation}
\label{qGaussian2}
with
\begin{equation}
N_q \equiv \left\{
\begin{array}{llll}
 \Bigl[ \dfrac{q-1}{\pi} \Bigr]^{1/2} \dfrac{\Gamma\Bigl(\dfrac{1}{q-1}\Bigr)}{\Gamma\Bigl(\dfrac{3-q}{2(q-1)}\Bigr)}  
&&&\textrm{if}\quad 1<q<3 \, , \\  
\\ 
 \dfrac{1}{\sqrt{\pi}}  &&&\textrm{if} \quad q=1 \,,      \\   
\\   
  \dfrac{3-q}{2}  \Bigl[\dfrac{1-q}{\pi} \Bigr]^{1/2} \dfrac{\Gamma\Bigl(\dfrac{3-q}{2(1-q)}\Bigr)}{\Gamma\Bigl(\dfrac{1}{1-q}\Bigr)}   
&&& \textrm{if} \quad q <1 \,,  
\end{array}\right.
\label{normqGaussian}
\end{equation}

We can verify that, for $q \ge 1$,
\begin{equation}
F_q[G_q(x)](\xi)=e_{\bar q}^{-\bar\beta\,\xi^2} \,,
\end{equation}
with
\begin{equation}
\bar q= \frac{1+q}{3-q} \,,
\label{eq1}
\end{equation}
and
\begin{equation}
\bar\beta= \frac{(3-q) N_q^{2(1-q)}}{8 \,\beta^{2-q}}\,.
\label{eq2}
\end{equation}
This expression can be conveniently rewritten as follows:
\begin{equation}
(\bar\beta)^{\frac{1}{\sqrt{2-q}}} \beta^{\sqrt{2-q}}= \Bigl[\frac{(3-q) N_q^{2(1-q)}}{8}\Bigr]^{\frac{1}{\sqrt{2-q}}}\,.
\end{equation}
For $q=1$, we recover $\bar\beta \beta=1/4$, which leads, in quantum mechanics, to the Heisenberg uncertainty principle. Notice also that Eqs. (\ref{eq1}) and (\ref{eq2}) are invertible, i.e., $(q,\beta) \to (\bar q, \bar \beta)$ and $(\bar q,\bar \beta) \to (q,\beta)$.

The above relations show that the $q$-FT is invertible within the class of the $q$-Gaussians. It is not so in general. Indeed, it has been shown by Hilhorst \cite{Hilhorst2009,Hilhorst2010} that, for a given value of $q$, one-parameter families of functions $\{f(x)\}$ exist such that their $q$-FT does {\it not} depend on that parameter. Therefore the $q$-FT has not always an unique pre-image, i.e., although the $q$-FT is invertible within the closed class of $q$-Gaussians, it is not invertible in general. This peculiar property does not exist for $q=1$, but it does emerge for $q>1$. Therefore an interesting question can be put: {\it Does a procedure exist which given a (say non-negative, for simplicity) function $f(x)$, enables us to go forward and backward through the $q$-FT and come back to the same function?} The answer is {\it yes}, as recently shown in \cite{JaureguiTsallis2010a}. The procedure is as follows. 

Given a non-negative function $f(x)$, we make the transformation $x \to x+y$, and then calculate the $q$-FT on the variable $x$, i.e., we calculate 
\begin{equation}
%\begin {align}
F_q[f_y](\xi,y)=\int_{-\infty}^{+\infty}f(x+y)e_q^{i \xi x[f(x+y)]^{q-1}}\,d x\,.
\label {qFTxy}
%\end {align}
\end{equation}
Next, by using the recently introduced $q$-Dirac delta \cite{JaureguiTsallis2010a}, we can recover the function $f(y)$, or equivalently $f(x)$, through 
\begin{equation}
f(y)=\left[\frac{2-q}{2\pi}\int_{-\infty}^{+\infty}F_q[f_y](\xi,y)\,d \xi\right]^{\frac{1}{2-q}}
\label{mainresult}
\end{equation}
for all points of the interior of the support of $f(x)$ (see \cite{JaureguiTsallis2010b} for details concerning the points at the edge of the support). 

\subsubsection{$q$-independence}

Two random variables $X$ and $Y$, with respective distributions $f_X(x)$ and $f_Y(y)$, are said {\it independent} if the joint distribution satisfies
\begin{equation}
f(x,y)=f_X(x)f_Y(y)
\end{equation}
This implies that the distribution of their sum is given by
\begin{equation}
f_{X+Y}(z)=\int dx \int dy f(x,y) \delta(z-x-y)=\int dx f_X(x)f_Y(z-x) \,, 
\end{equation}
where $\delta$ denotes the Dirac delta. If we take the FT we obtain that
\begin{equation}
F[f_{X+Y}(z)](\xi)=F[f_X(z)](\xi) \,F[f_Y(z)](\xi) \,. 
\end{equation}
Due to this property, we can alternatively define {\it independence} of $X$ and $Y$ by saying that the FT of $f_{X+Y}(z)$ equals the product of the FT of $f_X(x)$ and the FT of $f_Y(y)$. If the variables are {\it not} independent (i.e., if they are correlated), we have that
\begin{equation}
F[f_{X+Y}(z)](\xi)= \int dx \int dy f(x,y)\delta(z-x-y)         \ne F[f_X(z)](\xi) \,F[f_Y(z)](\xi) \,. 
\end{equation}

We shall next define a special type of correlation between $X$ and $Y$, named {\it $q$-independence}, by imposing
\begin{equation}
F_q[f_{X+Y}(z)](\xi)=F_q[f_X(z)](\xi) \otimes_{\bar q} F_q[f_Y(z)](\xi) \,, 
\end{equation}
with $\bar q$ given by Eq. (\ref{eq1}). This definition might seem rather esoteric at this stage, but it will turn out to be amazingly frequent in nature, as we shall see later on. Of course $1$-independence recovers independence.

\subsubsection{$q$-generalized central limit theorems}

In the empirical sciences, the repetition of an experiment (in nearly equal conditions) is a must if we wish to increase the experimental precision. If we do it $N$ times, we then consider as valid experimental value the {\it arithmetic mean} of those $N$ results. This ubiquitous fact points towards the importance of considering random variables of the type $S_N = X_1+X_2+...+X_N$, which constitutes the scope of the Central Limit Theorem (CLT). This important theorem admits a variety of forms, but basically it states that, if $\{X_i\}$ are equally distributed {\it independent} (or quasi-independent in some sense), have a {\it finite} variance, and $N$ is increasingly {\it large}, the sum  $S_N$ always converges, after appropriate centering and scaling, onto a Gaussian, which is therefore called the {\it attractor} of the sum. If the variance diverges, the attractor becomes instead a L\'evy distribution, also called $\alpha$-stable distribution with $\alpha<2$ (this is sometimes referred to as the L\'evy-Gnedenko CLT). 

It is quite natural that the attractors are neither Gaussian nor L\'evy distributions if strong correlations are present between the $\{X_i\}$ variables. It has been recently shown \cite{UmarovTsallisSteinberg2008} that, if the variables are equally distributed {\it $q$-independent} and a conveniently $q$-generalized variance $\sigma_Q$ is {\it finite}, the $N \to\infty$ attractors are $q$-Gaussians. If that $q$-generalized variance {\it diverges},  then the nature of the attractors is different and they are referred to as the $(q,\alpha)$-stable distributions $L_{q,\alpha}(x)$ \cite{UmarovTsallisGellMannSteinberg2010}. The $\alpha \to 2$ limit of the $(q,\alpha)$-stable distributions are the $q$-Gaussians $L_{q,2}(x) \equiv G_q(x)$; the $q \to 1$ limit of the $(q,\alpha)$-stable distributions are the $\alpha$-stable distributions $L_{1,\alpha}(x) \equiv L_\alpha(x)$; the limit $L_{1,2}(x) \equiv G(x)$ corresponds to Gaussians. $Q$ is defined as $Q=2q-1$ (with $1 \le q <2$), and $\sigma_Q = \frac{\int_{-\infty}^{\infty} dx\,x^2 \, [f(x)]^Q}{\int_{-\infty}^{\infty} dx\, [f(x)]^Q}$. These theorems are schematically depicted in the following Table (where the $C_{q,\alpha}$'s are positive coefficients). See \cite{TsallisQueiros2007,QueirosTsallis2007} for typical illustrations of the four types of attractors.

\begin{table}[htbp]
\begin{tabular}{c||c|c||}
                     &  $q=1$ [independent]                      & $q \ne 1$ (i.e., $Q \ne 1$) [globally correlated]                  \\
[1mm] \hline\hline
$\sigma_Q<\infty$      &  $G(x)$                                   & $G_q(x) $                                        \\   
   $(\alpha=2)$        & [with same $\sigma_1$ of $f(x)$]          & [with same $\sigma_Q$ of $f(x)$]                                   \\  
                     &                                           & $G_q(x) \sim G(x)$  \hspace{1.2cm} if $|x|<<x_c(q,2) $             \\
                     &                                           & $G_q(x) \sim C_{q,2}/|x|^{2/(q-1)}$  if $|x|>>x_c(q,2)$            \\
                     &                                           & \hspace{0.5cm} for $q>1$, with $\lim_{q \to 1} x_c(q,2) = \infty$  \\
[3mm] \hline
$\sigma_Q \to \infty$  &  $L_\alpha(x)$                                                   &$L_{q,\alpha}(x)$                                  \\  
 $(\alpha < 2)$        &[with same $|x| \to \infty$ behavior of $f(x)$]                   &[with same $|x| \to \infty$ behavior of $f(x)$]   \\ 
                  &$L_{\alpha}(x) \sim G(x)$ \hspace{0.8cm} if $|x|<< x_c(1,\alpha)$ & $L_{q,\alpha} \sim    C_{q,\alpha}^{(intermediate)}/|x|^{\frac{2(1-q)-\alpha(3-q)}{2(q-1)}}$               \\
                  &$L_{\alpha}(x) \sim C_{1,\alpha}/|x|^{1+\alpha}$  if $|x|>> x_c(1,\alpha)$   &  \hspace{0.9cm} if $ x_c^{(1)}(q,\alpha)<< |x|< < x_c^{(2)}(q,\alpha)$    \\
                  & with $\lim_{\alpha \to 2} x_c(1,\alpha) =\infty$                            &   $L_{q,\alpha} \sim C_{q,\alpha}^{(distant)}/|x|^{\frac{1+\alpha}{1+\alpha (q-1)}}$ \hspace{1.2cm}                           \\
                  &                                                                             & \hspace{-1.0cm} if $ |x|>> x_c^{(2)}(q,\alpha)$  \\
[3mm] \hline \hline
\end{tabular}
\end{table}

There are many Gaussians in nature: it seems reasonable to believe that this is due to the CLT, which shows that details can be of no importance if the number of involved random variables is large. Similarly, and for the same reason, we also expect many $q$-Gaussians to exist in nature (as well as in artificial and social systems), as long as $q$-independence emerges naturally in many complex systems, where strong correlations between elements is an important ingredient. And it is precisely this situation that has been profusely illustrated in Section 1.   

Let us incidentally mention that the lack of inverse of the $q$-FT has made Hilhorst \cite{Hilhorst2009,Hilhorst2010} to disregard $q$-Gaussians as attractors. A detailed reply to his claim has been recently made available in \cite{UmarovTsallis2010}, which reinforces that $q$-Gaussians are, in many respects, very special distributions. It must be also taken into account that, from a different perspective, a theorem like the $q$-CLT (and even other forms associated with other types of correlations) has been also proved {\it without} using the $q$-FT \cite{VignatPlastino2007,HahnJiangUmarov2010}.

The above discussion mainly focused $q \ge 1$. However, the $q$-FT has also been addressed for $q<1$ (see \cite{NelsonUmarov2008,NelsonUmarov2010}).

As an interesting mathematical challenge we may now ask: {\it Is it possible to alternatively prove the $q$-CLT by simultaneously using the $q$-FT and relation (\ref{mainresult})?}  

\subsubsection{Possible relation between $q$-independence and scale-invariance}

Two probabilistic models, \cite{MoyanoTsallisGellMann2006} and \cite{ThistletonMarshNelsonTsallis2009} respectively, involving $N$ equally distributed random variables  were introduced some time ago. Their numerical discussion suggested that, in the $N \to\infty$ limit, $q$-Gaussians emerged with $q \le 1$, after appropriate centering and scaling. It was however proved \cite{HilhorstSchehr2007} quickly after that the limiting distributions of these two models are {\it not} exactly $q$-Gaussians, even if numerically they are amazingly close to them. This interesting result put forward a relevant question, which we describe now.  

We will consider {\it scale-invariant} a $N$-particle probabilistic model which satisfies the following property:
\begin{equation}
f_{N-1}(x_1,x_2,...,x_{N-1})=\int dx_N \,f_{N}(x_1,x_2,...,x_{N})\,,
\end{equation}
where $f_{N}(x_1,x_2,...,x_{N})$ is the joint probability distribution associated with the random variables $(x_1,x_2,...,x_N)$, satisfying $\int dx_1\,dx_2...dx_N \,f_{N}(x_1,x_2,...,x_{N})=1\;(\forall N)$. In other words, a $N$-particle probabilistic model is said scale invariant if the marginal probabilities after tracing over any particle the $N$-particle joint probabilities coincide with the $(N-1)$-particle joint probabilities. If the system is made by $N$ binary random variables, scale invariance is nothing but what is sometimes referred to as the {\it Leibnitz triangle rule} (not to be confused with the {\it Leibnitz chain rule}!), i.e. (see for instance \cite{TsallisGellMannSato2005}),
\begin{equation}
r_{N,n-1}+r_{N,n}=r_{N-1,n} \;\;(N=2,3,4,...;\,n=0,1,2,...,N) \,,
\end{equation}
with
\begin{equation}
\sum_{n=0}^N \frac{N!}{(N-n)\,!n!}\,r_{N,n}=1 \;\;(\forall N) \,.
\end{equation}

It is straightforward to prove that probabilistic independence yields scale-invariance. Indeed, if no correlation is present, we have that $f_{N}(x_1,x_2,...,x_{N})=f_1(x_1)\,f_1(x_2)\,...,f_1(x_N)$ with $\int dx\,f_1(x)=1$, which trivially implies scale-invariance. But, even in the presence of correlations, scale-invariance is possible. The celebrated Leibnitz triangle is one such example. Both models introduced in  \cite{MoyanoTsallisGellMann2006,ThistletonMarshNelsonTsallis2009} also are nontrivially scale-invariant. But they are not $q$-independent. Indeed, if they were, the $N\to\infty$ limiting distributions of both models would be $q$-Gaussians, and they are not. 

We immediately conclude that scale invariance is not sufficient for $q$-independence. It is nevertheless compatible with it. Indeed, in \cite{RodriguezSchwammleTsallis2008,HanelThurnerTsallis2009} we have introduced scale-invariant models whose limiting distributions are $q$-Gaussians for values of $q$ both above and below $q=1$ (which corresponds in fact to independence).     
                         
The present scenario is as follows: scale-invariance clearly is not sufficient for $q$-independence, but could well be necessary. 

So the following question emerges: {\it What are the exact mathematical implications between (strict or asymptotic) scale-invariance and (strict or quasi) $q$-independence? Or, alternatively, is there some other generic condition which, added to scale-invariance, makes it necessary and sufficient for $q$-independence?}                          
                         
\subsection{q-generalized plane waves, Dirac delta, wave equation and nonlinear Klein-Gordon, Schroedinger and Dirac equations}
         
The $q$-exponential function has various remarkable properties which straightforwardly generalize those of the exponential function. Let us illustrate this through the following ordinary differential equations (ODE).
We consider the (linear) ODE
\begin{equation}
\frac{dy}{dx}=a_1 \,y \;\;\;(a_1 \in {\cal R}) \,,
\label{linear}
\end{equation}
with $y(0)=1$. The solution is given by
\begin{equation}
y(x)=e^{\,a_1\,x} \,.
\end{equation}
We may now generalize Eq. (\ref{linear}) into the following (nonlinear) one:
\begin{equation}
\frac{dy}{dx}=a_q \,y^q \;\;\;((a_q, q) \in {\cal R}^2)\,.
\label{nonlinear}
\end{equation}
The solution is now given by
\begin{equation}
y(x)=e_q^{\,a_q\,x} \,.
\end{equation}
We may finally unify \cite{TsallisBemskiMendes1999} both Eqs. (\ref{linear}) and (\ref{nonlinear}) into
\begin{equation}
\frac{dy}{dx}=a_1 \, y + (a_q-a_1) \,y^q \;\;\;((a_1,a_q, q) \in {\cal R}^3)\,,
\label{linearnonlinear}
\end{equation}
whose solution is given by
\begin{equation}
y= \Bigl[1- \frac{a_q}{a_1} + \frac{a_q}{a_1}\,e^{(1-q) \,a_1 \,x} \Bigr]^{\frac{1}{1-q}} \,.    
\label{6.1b}
\end{equation}
This solution makes a crossover from the $q$-exponential (for small $x$) to the exponential function (for $|(1-q) \,a_1 \,x| >>1$). As an interesting remark, let us mention that, for $q=2$, this solution becomes y= $\frac{1}{1- \frac{a_q}{a_1} + \frac{a_q}{a_1}\,e^{- \,a_1 \,x}}$, which in the appropriate limits (and with the photonic density of states) recovers the Planck law for the black-body radiation! (see details in \cite{Tsallis2009a}). 

Let us now turn onto $q$-plane waves and related matters, and later on we will address a possible relation between them and the crossover we have mentioned here above.  

To start we remind that the representation of Dirac delta in terms of plane waves has been recently generalized as follows \cite{JaureguiTsallis2010b} in terms of $q$-plane waves:
\begin{equation}
\delta(x) = \frac{2-q}{2\pi} \int_{-\infty}^{\infty}dk\, e_q^{-ikx} \;\;\;\;(1 \le q<2) \,,
\end{equation}
which implies that a wide class of functions $f(x)$ exist such that
\begin{equation}
\int_{-\infty}^\infty dx\,\delta(x-x_0) f(x) = f(x_0)\;\;\;\;(1 \le q<2) \,.
\label{qdelta}
\end{equation}

Let us now focus on the standard one-dimensional linear wave equation
\begin{equation}
\label{eq:1dlwaveeq}
{\partial ^{2} \Phi(x,t) \over \partial x^{2}} = 
{1 \over c^{2}}
{\partial ^{2} \Phi(x,t) \over \partial t^{2}}~,   
\end{equation}
for which any function of the type $\Phi(kx-\omega t)$, twice differentiable,
is a solution. In particular, one may have a {\it $q$-plane wave},   
\begin{equation}
\label{eq:qsolwaveeq}
\Phi(x,t) = \Phi_{0} \, \exp_{q} \left[ i (kx-\omega t) \right]~; \quad 
[\Phi_{0} \equiv \Phi(0,0)]~, 
\end{equation}
as a solution of the equation above, provided that \cite{NobreMonteiroTsallis2010}
\begin{equation}
\omega = c \, k \;\;\;\;(1 \le q<3) \,.
\end{equation} 
Its generalization to $d$ dimensions is straightforward (see \cite{NobreMonteiroTsallis2010}). Its plane wave solution becomes

\begin{equation}
\label{3dqsolwaveeq}
\Phi(\vec{x},t) = \Phi_{0} \, \exp_{q} \left[ i (\vec{k} \cdot \vec{x}
-\omega t) \right]~. 
\end{equation}

Let us introduce now the following $d$-dimensional 
nonlinear generalization of
the Schr\"odinger equation for a free particle of mass $m$, 
\begin{equation}
\label{eq:schreq}
i \hbar {\partial \over \partial t} 
\left[ \frac{\Phi(\vec{x},t)}{\Phi_{0}} \right] 
= - {1 \over 2-q} \ \frac{\hbar ^{2}}{2m}
\nabla^{2} \left[ \frac{\Phi(\vec{x},t)}{\Phi_{0}} \right]^{2-q}~.  
\end{equation}
We notice that the scaling of the wave function by 
$\Phi_{0}$ guarantees the correct physical dimensionalities for all terms.
This scaling becomes 
irrelevant only for linear equations [e.g., 
in the particular case $q=1$ of Eq. (\ref{eq:schreq}). 
Now, if ones uses the $q$-plane wave solution  Eq. (\ref{3dqsolwaveeq}) by simply replacing 
$\vec{k} \rightarrow \vec{p}/\hbar$ and  
$\omega \rightarrow E/\hbar$, one verifies \cite{NobreMonteiroTsallis2010} that the form 
\begin{equation}
\label{qplane}
\Phi(\vec{x},t) = \Phi_{0} \, \exp_{q} \left[\frac{i}{\hbar} (\vec{p} \cdot \vec{x}
-E t) \right]~, 
\end{equation}
is a solution of the above nonlinear equation, with 
\begin{equation}
E=p^{2}/2m \,, 
\end{equation}
thus preserving the well known energy spectrum of the free particle {\it for all values of $q$}.

Let us now add a mass term in the traditional wave equation, and propose the following nonlinear Klein-Gordon equation in $d$ dimensions, namely 
\begin{equation}
\label{eq:kgordoneq}
\nabla^{2} \Phi(\vec{x},t) = 
{1 \over c^{2}}
{\partial ^{2} \Phi(\vec{x},t) \over \partial t^{2}} + 
q \, {m^{2}c^{2} \over \hbar^{2}} \, \Phi(\vec{x},t) \left[
{\Phi(\vec{x},t) \over \Phi_{0}} \right]^{2(q-1)}~.  
\end{equation}
One may verify easily \cite{NobreMonteiroTsallis2010} that the same $q$-plane wave used for the nonlinear Schr\"odinger 
equation is a solution of Eq. (\ref{eq:kgordoneq}), {\it preserving for all $q$ the Einstein relation}  
\begin{equation}
\label{eq:einsteinrel}
E^{2} = p^{2}c^{2} + m^{2}c^{4}~. 
\end{equation}

Let us finally consider a nonlinear generalization of the $d=3$ Dirac equation \cite{Dirac1928}, namely \cite{NobreMonteiroTsallis2010}
\begin{equation}
\label{eq:diraceq}
i \hbar \ {\partial \Phi(\vec{x},t) \over \partial t}
+ i \hbar c (\vec{\alpha} \cdot \vec{\nabla}) \Phi(\vec{x},t)
= \beta mc^{2} A^{(q)}(\vec{x},t) \ \Phi(\vec{x},t)~, 
\end{equation}
where $\alpha_{x},\alpha_{y},\alpha_{z}$ (written in terms of the Pauli
spin $2 \times 2$ matrices $\sigma_x$, $\sigma_y$, and $\sigma_z$) and $\beta$ (written in terms of the $2 \times 2$ identity
matrix $I$) are the standard $4 \times 4$ matrices~\cite{Liboff2003}, namely

\begin{equation}
\alpha_x=
\left(
\begin{array}{cc}
0 & \sigma_x\\
\sigma_x & 0
\end{array} 
\right), \;\;\;
\alpha_y=
\left(
\begin{array}{cc}
0 & \sigma_y\\
\sigma_y & 0
\end{array} 
\right), \;\;\;
\alpha_z=
\left(
\begin{array}{cc}
0 & \sigma_y\\
\sigma_y & 0
\end{array} 
\right),
\end{equation}
and
\begin{equation}
\beta=
\left(
\begin{array}{cc}
I & 0\\
0 & -I
\end{array} 
\right).
\end{equation}

The new, $q$-dependent, term is given by the $4 \times 4$ diagonal matrix
$A^{(q)}_{ij}(\vec{x},t) = \delta_{ij}[\Phi_{j}(\vec{x},t)/a_{j}]^{q-1}$,
where $\{a_{j}\}$ are complex constants
($A^{(1)}_{ij}(\vec{x},t)=\delta_{ij}$).  
The solution of Eq. (\ref{eq:diraceq}) we focus on is the following
four-component column matrix 
\begin{equation}
\label{eq:quadsol}
\Phi(\vec{x},t) \equiv 
\left( 
\begin{array}{c}
\Phi_{1}(\vec{x},t) \\
\Phi_{2}(\vec{x},t) \\
\Phi_{3}(\vec{x},t) \\
\Phi_{4}(\vec{x},t)
\end{array}
\right) 
= 
\left( 
\begin{array}{c}
a_{1} \\
a_{2} \\
a_{3} \\
a_{4}
\end{array}
\right) 
\, \exp_{q} \left[ {i \over \hbar} 
(\vec{p} \cdot \vec{x} - Et) \right]~.
\end{equation}
Substituting this four-component vector 
into Eq.(\ref{eq:diraceq}), we get, for the 
coefficients $\{ a_{j} \}$, {\it precisely the same set of four algebraic
equations  corresponding to the linear case}, namely (see page 803, Eq. (15.45b) of
\cite{Liboff2003}), 
\begin{eqnarray}
(E-mc^2)a_1 - cp_za_3-c(p_x-ip_y)a_4=0 \nonumber \\
(E-mc^2)a_2 - c(p_x +ip_y)a_3+cp_za_4=0 \nonumber \\
(E-mc^2)a_3 - cp_za_1-c(p_x-ip_y)a_2=0  \nonumber \\
(E-mc^2)a_4 - c(p_x+ip_y)a_1+cp_za_2=0 
\end{eqnarray}

These equations have, for all $q$, a nontrivial solution only if the Einstein energy-momentum relation Eq. (\ref{eq:einsteinrel}) is satisfied.   
 
The above nonlinear Schroedinger equation (\ref{eq:schreq}) can be further generalized as follows:  
\begin{equation}
\label{furthereq:schreq}
i \hbar {\partial \over \partial t} 
\left[ \frac{\Phi(\vec{x},t)}{\Phi_{0}} \right] 
= -  \ \frac{\hbar ^{2}}{2m} \Bigl\{a_1^S \, \nabla^{2} \left[ \frac{\Phi(\vec{x},t)}{\Phi_{0}} \right] + (a_q^S-a_1^S) {1 \over 2-q}
\nabla^{2} \left[ \frac{\Phi(\vec{x},t)}{\Phi_{0}} \right]^{2-q}\Bigr\}~.  
\end{equation}  
The case $(a_1^S,a_q^S)=(1,1)$ (or equivalently $(a_q^S,q)=(1,1)$) recovers the usual linear Schroedinger equation, and the case  $(a_1^S,a_q^S)=(0,1)$ recovers Eq. (\ref{eq:schreq}).

Analogously, the nonlinear Klein-Gordon equation (\ref{eq:kgordoneq}) can be further generalized as follows:
\begin{equation}
\label{furthereq:kgordoneq}
\nabla^{2} \Phi(\vec{x},t) = 
{1 \over c^{2}}
{\partial ^{2} \Phi(\vec{x},t) \over \partial t^{2}} + 
{m^{2}c^{2} \over \hbar^{2}} \, \Phi(\vec{x},t) \Bigl\{a_1^{KG} + (a_q^{KG}-a_1^{KG}) \, q\left[
{\Phi(\vec{x},t) \over \Phi_{0}} \right]^{2(q-1)} \Bigr\}.  
\end{equation}
The case $(a_1^{KG},a_q^{KG})=(1,1)$ (or equivalently $(a_q^{KG},q)=(1,1)$) recovers the usual linear Klein-Gordon equation, and the case  $(a_1^{KG},a_q^{KG})=(0,1)$ recovers Eq. (\ref{eq:kgordoneq}).

Finally, the nonlinear Dirac equation (\ref{eq:diraceq}) can be further generalized as follows:
\begin{equation}
\label{furthereq:diraceq}
i \hbar \ {\partial \Phi(\vec{x},t) \over \partial t}
+ i \hbar c (\vec{\alpha} \cdot \vec{\nabla}) \Phi(\vec{x},t)
= \beta mc^{2} \delta_{ij} \Bigl\{a_1^D +(a_q^D-a_1^D)  [\Phi_{j}(\vec{x},t)/a_{j}]^{q-1}    \Bigr\} \ \Phi(\vec{x},t). 
\end{equation} 
The case $(a_1^{D},a_q^{D})=(1,1)$ (or equivalently $(a_q^D,q)=(1,1)$) recovers the usual linear Klein-Gordon equation, and the case  $(a_1^{D},a_q^{D})=(0,1)$ recovers Eq. (\ref{eq:diraceq}).

At the light of the situations that have been considered in the present Subsection, we may ask the following two points:
{\it What is the precise class of functions $f(x)$ for which Eq. (\ref{qdelta}) applies?} (See \cite{JaureguiTsallis2010a}) {\it Is it possible, following along the lines of Eqs (\ref{linearnonlinear}) and (\ref{6.1b}) or any other path, to find exact solutions of Eqs. (\ref{furthereq:schreq}), (\ref{furthereq:kgordoneq}) and (\ref{furthereq:diraceq})?}

\subsection{Dependence of q on the interaction-range of many-body Hamiltonians} \label{manybody}
                       
The BG entropy and its associated statistical mechanics are known to be very useful for many-body Hamiltonians with say two-body interactions that are short-ranged and that do not introduce severe frustration (which would break down ergodicity, like it happens for instance for spin-glasses). A paradigmatic classical Hamiltonian system which violates (for $0 \le \alpha \le d$) the short-range-interacting condition (and introduces no frustration at all) is the $\alpha$-XY one \cite{AnteneodoTsallis1998} \footnote{Many other types of classical and quantum Hamiltonian systems can be thought of in long-range-interacting versions as illustrations along similar lines (see, for instance, \cite{NobreTsallis1995,CarideTsallisZanette1983}).}:

\begin{equation}
{\cal H}_N= \frac{1}{2}\sum_{i=1}^N p_i^2 + \frac{1}{2 \tilde N}\sum_{i,j} \frac{1-cos (\theta_i - \theta_j)}{r_{i,j}^\alpha} \;\;\;(\alpha \ge 0)\,,
\label{hamiltonian}
\end{equation}
where the $N$ rotators are located at the sites of a $d$-dimensional simple hypercubic lattice (with periodic boundary conditions). The distances $r_{i,j}$ take the values $1,2,3,...$ for $d=1$, the values  $1,\sqrt{2},2, ...$ for $d=2$, the values $1,\sqrt{2},\sqrt{3},2,...$ for $d=3$, and so on; due to the periodic boundary conditions, more than one value of $r_{i,j}$ can be defined between two given sites $i$ and $j$ (e.g., for $d=1$, two such values exist generically): in all cases we take into consideration only the smallest value of $r_{i,j}$ for a given couple $(i,j)$. $\tilde N$ is defined as follows:
\begin{equation}
\tilde N \equiv \sum_{j} \frac{1}{r_{i,j}^\alpha}\,.
\end{equation}
The role played by $\tilde N$ is analyzed in \cite{AnteneodoTsallis1998}. It is introduced here to conform to the vast literature existing for this Hamiltonian. It makes the Hamiltonian ${\cal H}_N$ to be extensive for all admissible values of $\alpha$.

The (infinitely degenerate) fundamental state of the Hamiltonian (\ref{hamiltonian}) corresponds to all rotators being parallel, and the corresponding total energy $U_N$ vanishes. We may define the asymptotic energy per particle $u \equiv \lim_{N\to\infty} U_N/N$. A critical value $u_c$ exists such that the system is ferromagnetically ordered for $0 \le u<u_c$, and it is paramagnetically disordered for $u \ge u_c$. For example, for $d=1$, it is $u_c=3/4$, $\forall \alpha$ (see \cite{CampaGiansantiMoroniTsallis2001,Tsallis2009a} for further details). 

For all values $\alpha/d>1$, only one collective thermal equilibrium exists, correctly (and analytically) described within BG statistical mechanics. For those systems, the limits $\lim_{N \to\infty} \lim_{t \to\infty}$ and $\lim_{t \to\infty} \lim_{N \to\infty}$ of all thermostatistical properties commute.  For all values $0 \le \alpha/d<1$, two collective thermal stationary states exist, namely the {\it thermal equilibrium} (corresponding to the ordering first $t \to\infty$ and then $N \to\infty$), and the so-called {\it quasi-stationary state} (QSS) (corresponding to the ordering first $N \to\infty$ and then $t \to\infty$). Thermal equilibrium is correctly described within the BG theory. Not so the QSS, which emerge at values of $u$ slightly below $u_c$ (e.g., for $u \simeq 0.69$ for $d=1$), and are {\it not} described correctly within the BG theory. Indeed, for QSS, the microscopic dynamics is {\it nonergodic} since the ensemble averages and the time averages do {\it not} coincide. The ensemble averages are neither BG ($q=1$) nor $q \ne 1$ well described (approaches based on the Vlasov equation might be relevant). But the time averages (by far the most relevant in experiments)  appear to follow $q$-statistics \cite{PluchinoRapisardaTsallis2007,PluchinoRapisardaTsallis2008}. Indeed, the distribution of momenta $\{p_i\}$ is definitively non Gaussian, and is well fitted by $q_{vel}$-Gaussians with $q_{vel}>1$, where $vel$ stands for $velocities$. For the $\alpha = 0$ model (known in the literature as the HMF model \cite{AntoniRuffo1995}), we have $q_{vel} \simeq 1.5$ \cite{PluchinoRapisardaTsallis2007,PluchinoRapisardaTsallis2008}. In principle one expects $q_{vel}(\alpha,d)$ with  $q_{vel}(0,d)$ being independent from $d$. It is however, quite probable that $q_{vel}$ only depends on the ratio $\alpha/d$, i.e., $q_{vel}(\alpha/d)$ with $q_{vel}(0) \simeq 1.5$. Moreover, it seems plausible that $q_{vel}(\alpha/d)$ monotonically decreases from $q_{vel}(0)$ to $q_{vel}(1)=1$ when $\alpha/d$ increases from zero to unity. 

A directly related question is the following. It is clear that, at the QSS, the corresponding (canonical) distribution for the entire phase space can {\it not} be given by the BG weight, i.e., we have that 
\begin{equation}
P(p_1,p_2,...,p_N, \theta_1,\theta_2,...,\theta_N) \ne \frac{e^{-\beta \, {\cal H}_N(p_1,p_2,...,p_N,\theta_1,\theta_2,...,\theta_N)}}{\int dp_1 dp_2...dp_N d\theta_1 d\theta_2...d\theta_N \, e^{-\beta \, {\cal H}_N(p_1,p_2,...,p_N,\theta_1,\theta_2,...,\theta_N)}} \,. 
\end{equation}
Indeed, if the BG weight was the correct one, we would have (since the kinetic and potential energies commute) that the one-momentum marginal distribution $P(p_1)$ would be a Gaussian, namely the  Maxwellian $\frac{e^{-\beta \, p_1^2/2}}{\int dp_1 \, e^{-\beta \, p_1^2/2)}}$. But we numerically know that this is not true. Therefore the distribution $P(p_1,p_2,...,p_N, \theta_1,\theta_2,...,\theta_N)$ must be a different one. A possible Ansatz would be that, for $N >>1$, it is given by
\begin{equation}
P(p_1,p_2,...,p_N, \theta_1,\theta_2,...,\theta_N) = \frac{e_{q_{stationary\,state}}^{-\beta_{q_{stationary\,state}} \, {\cal H}_N(p_1,p_2,...,p_N,\theta_1,\theta_2,...,\theta_N)}}{\int dp_1 dp_2...dp_N d\theta_1 d\theta_2...d\theta_N \, e_{q_{stationary\,state}}^{-\beta_{q_{stationary\,state}} \, {\cal H}_N(p_1,p_2,...,p_N,\theta_1,\theta_2,...,\theta_N)}} \,, 
\label{qexponential}
\end{equation}
where
\begin{equation}
q_{stationary\,state}=f(\alpha/d) \,,
\end{equation}                    
and
\begin{equation}
\beta_{q_{stationary \, state}}= g(\alpha/d,\beta) \,.
\end{equation}    
Naturally we expect $f(\alpha/d)=1$ and $g(\alpha/d,\beta)=\beta$ for $\alpha/d>1$. 

At this point, relevant open questions that arise are: {\it Is it true that, at the QSS, the distribution in the full $\Gamma$ phase space is given by Eq.  (\ref{qexponential})?} {\it If so, what is the function $q_{stationary\,state}=f(\alpha/d)$ for $0 \le \alpha/d <1$?} {\it Is it true that the one-moment marginal distribution  is a $q_{vel}$-Gaussian?} {\it Is it $q_{vel}=q_{stationary \, state}$, or, if not, what is it their relationship for $0 \le \alpha/d <1$?} {\it Are $q_{vel}(\alpha/d)$ and $q_{stationary\,state}(\alpha/d)$ universal in the sense that they are shared by a wide class of classical Hamiltonian systems?}

\subsection{Temperature and the zeroth principle of thermodynamics}

\begin{figure}
\begin{center}
\includegraphics[width=13.5 cm,angle=0]{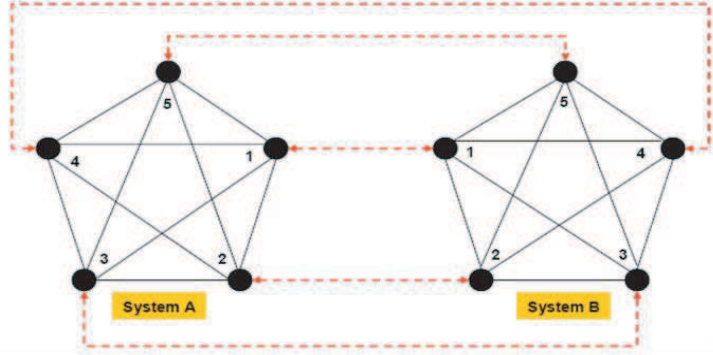}
\end{center}
\vspace{-1.5cm}
\caption{\small Systems $A$ and $B$ that will be put in thermal contact at a certain moment by allowing the coupling constant $l$ to become different from zero. 
Here $N=5$. From \cite{Albuquerque2008}.
}
\label{Albuquerque2008a}
\end{figure}

\begin{figure}
\begin{center}
\includegraphics[width=13.5 cm,angle=0]{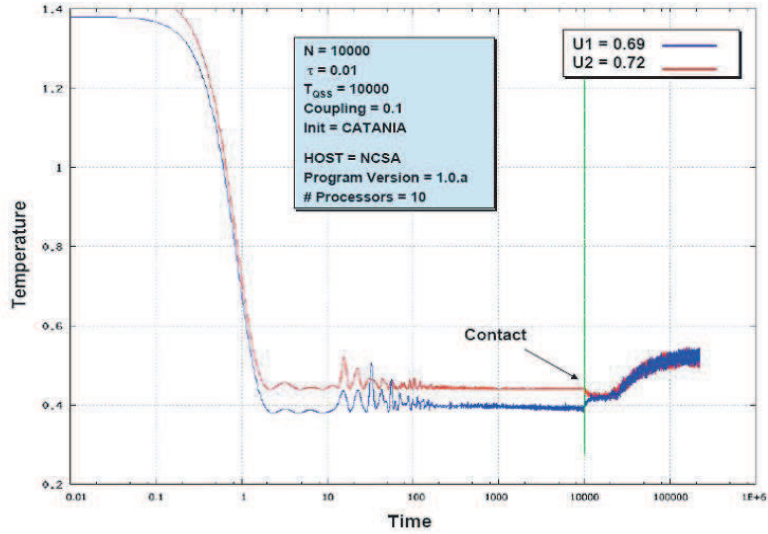}
\end{center}
\vspace{-1.5cm}
\caption{\small Time evolution of the temperatures of $A$ and $B$. The initial conditions are water bag for both $A$ and $B$, at slightly different initial internal energies, hence slightly different initial temperatures. Here $N=10000$, and $l$ is taken zero until the moment indicated with a green vertical line, and $l=0.1$ after that moment. From \cite{Albuquerque2008}.
}
\label{Albuquerque2008b}
\end{figure}

The zeroth principle of thermodynamics plays a basic role in its axiomatic formulation. It states that if a system A is in thermal equilibrium with a system B, and B is in thermal equilibrium with a system C, then A is in thermal equilibrium with C. In other words, the concept of thermal equilibrium is transitive, which exhibits the great importance of the temperature, a quantity whose value is shared by all systems in thermal equilibrium. Can stationary (or quasi-stationary) states different from thermal equlibrium also satisfy this transitivity? It might be that some of them can. Indeed, in what concerns the behavior of two systems A and B at in QSS at somewhat different temperatures and being put in thermal contact suggests that. After contact between two equally sized systems (with $N$ rotators and $N(N-1)/2$ links each), they evolve into a single double-sized system (with $2N$ rotators) whose temperature is between the two initial temperatures: see Figs. \ref{Albuquerque2008a} and \ref{Albuquerque2008b}, and details in \cite{Tsallis2009a}. All this occurs {\it before} the entire system makes the crossover to the BG regime, where thermal equilibrium takes place.    

The point which remains to be checked (for this and other similar long-ranged Hamiltonians) is: {\it If we follow a full sequence of connections and disconnections between systems A, B and C, all of them at their respective QSS's, will the behavior be totally analogous to what is known to happen at thermal equilibrium (i.e., at the BG states)?}

\subsection{Connection to thermodynamics and $q$-expectation values}

To generalize BG statistical mechanics for the canonical ensemble (from \cite{Tsallis2009b}), we optimize $S_q$ with the constraint 
\begin{equation}
\sum_{i=1}^W p_i=1
\end{equation}
and also
\begin{equation}
\sum_{i=1}^WP_iE_i = U_q \,,
\label{escortconstraint}
\end{equation}
where 
\begin{equation}
P_i \equiv \frac{p_i^q}{\sum_{j=1}^W p_i^q} \;\;\;\Bigl(\sum_{i=1}^WP_i=1 \Bigr)
\end{equation}
is the so-called {\it escort distribution} \cite{BeckSchlogl1993}. It follows that $p_i=\frac{P_i^{1/q}}{\sum_{j=1}^W P_j^{1/q}}$ . There are various converging reasons for being appropriate to impose the energy constraint with the $\{P_i\}$ instead of with the original $\{p_i\}$. The full discussion of this delicate point is beyond the present scope. However, some of these intertwined reasons are explored in \cite{Tsallis2004a}. By imposing Eq. (\ref{escortconstraint}), we follow \cite{TsallisMendesPlastino1998}, which in turn reformulates the results presented in \cite{Tsallis1988,CuradoTsallis1991}. The passage from one to the other of the various existing formulations of the above optimization problem are discussed in detail in \cite{TsallisMendesPlastino1998,FerriMartinezPlastino2005}.

The entropy optimization yields, for the stationary state,

\begin{equation}
p_i =\frac{e_{q}^{- \beta_q (E_i-U_q)}}{\bar{Z}_q} \,,  
\label{3.6.9c}
\end{equation}
with
\begin{equation}
\beta_q \equiv \frac{\beta}{\sum_{j=1}^W p_j^q} \,,
\label{3.6.9d}
\end{equation}
and
\begin{equation}
\bar{Z}_q \equiv \sum_i^W e_{q}^{- \beta_q (E_i-U_q)} \,,
\label{3.6.9e}
\end{equation}
$\beta$ being the Lagrange parameter associated with the constraint (\ref{escortconstraint}). Eq. (\ref{3.6.9c}) makes explicit that the probability distribution is, for fixed $\beta_q$, invariant with regard to the arbitrary choice of the zero of energies. 
The stationary state (or (meta)equilibrium) distribution (\ref{3.6.9c}) can be rewritten as follows: 
\begin{equation}
p_i= \frac{ e_q^{-\beta_q^\prime E_i}     }{Z_q^\prime} \;,
\label{3.6.9f}
\end{equation}
with
\begin{equation}
Z_q^\prime \equiv \sum_{j=1}^W e_q^{-\beta_q^\prime E_j} \;,
\label{3.6.9ff}
\end{equation}
and
\begin{equation}
\beta_q^\prime \equiv \frac{\beta_q}{1+(1-q) \beta_qU_q}\;.
\end{equation}
The form (\ref{3.6.9f}) is particularly convenient for many applications where comparison with experimental or computational data is involved. Also, it makes clear that $p_i$ asymptotically decays like $1/E_i^{1/(q-1)}$ for $q>1$, and has a cutoff for $q<1$, instead of the exponential decay with $E_i$ for $q=1$.

The connection to thermodynamics is established in what follows. It can be proved that
\begin{equation}
\frac{1}{T}=\frac{\partial S_q}{\partial U_q}\;,
\end{equation}
with $T \equiv 1/(k\beta)$. Also we prove, for the free energy,  
\begin{equation}
F_q \equiv U_q-TS_q= -\frac{1}{\beta} \ln_q Z_q\;,
\end{equation}
where
\begin{equation}
\ln_q Z_q = \ln_q {\bar Z}_q - \beta U_q\;.
\end{equation}
This relation takes into account the trivial fact that, in contrast with what is usually done in BG statistics, the energies $\{E_i\}$ are here referred to $U_q$ in (\ref{3.6.9c}). It can also be proved
\begin{equation}
U_q =-\frac{\partial}{\partial \beta} \ln_q Z_q \;,
\end{equation}
as well as relations such as
\begin{equation}
C_q \equiv T\frac{\partial S_q}{\partial T} = \frac{\partial U_q}{\partial T} = -T \frac{\partial^2 F_q}{\partial T^2} \;.
\end{equation}
In fact, the entire Legendre transformation structure of thermodynamics is $q$-invariant, which is both remarkable and welcome.

Important questions that remain to be clarified include: {\it What are the connections of quantities such as $\beta$, $\beta_q$, $\beta_q^\prime$, and $U_q$,  with the thermostatistical quantities naturally appearing in models such as those focused on in Subsection \ref{manybody}?}

\subsection{q-triplet and possibly associated algebras}

For the most basic quantities (e.g., sensitivity to the initial conditions, relaxation towards equilibrium of correlation functions, equilibrium distribution of energies) of systems described by the BG theory, the exponential function emerges ubiquitously as the adequate one. This function is replaced by the $q$-exponential one for systems described by nonextensive statistical mechanics. The question appears about what is the value of $q$ to be used.  A wide number of examples show that the value of $q$ is directly associated with the class of properties that are being studied. For example, for dissipative one-dimensional maps at the edge of chaos we have that the sensitivity to the initial conditions is characterized by $q_{sensitivity} <1$, the entropy production is characterized by $q_{entropy \, production}$ (and for such systems we verify that $q_{sensitivity}=q_{entropy \, production}$), the relaxation of the entropy towards its saturation value is characterized by $q_{relaxation}>1$, the sums of many successive iterations are characterized by $q_{attractor}>1$, and so on. In other words, for a given nonextensive system, we typically have {\it not one} value of $q$, but an {\it infinite number} of them, most probably interconnected in such a way that at the end only one (or very few) are independent, thus characterizing universality classes of nonextensivity. In the limit when a BG regime is approached, all these values of $q$ are typically expected to merge into the single value $q=1$.     

\begin{figure}[h]
\begin{center}
\hspace{-0.5cm}
\psfig{file=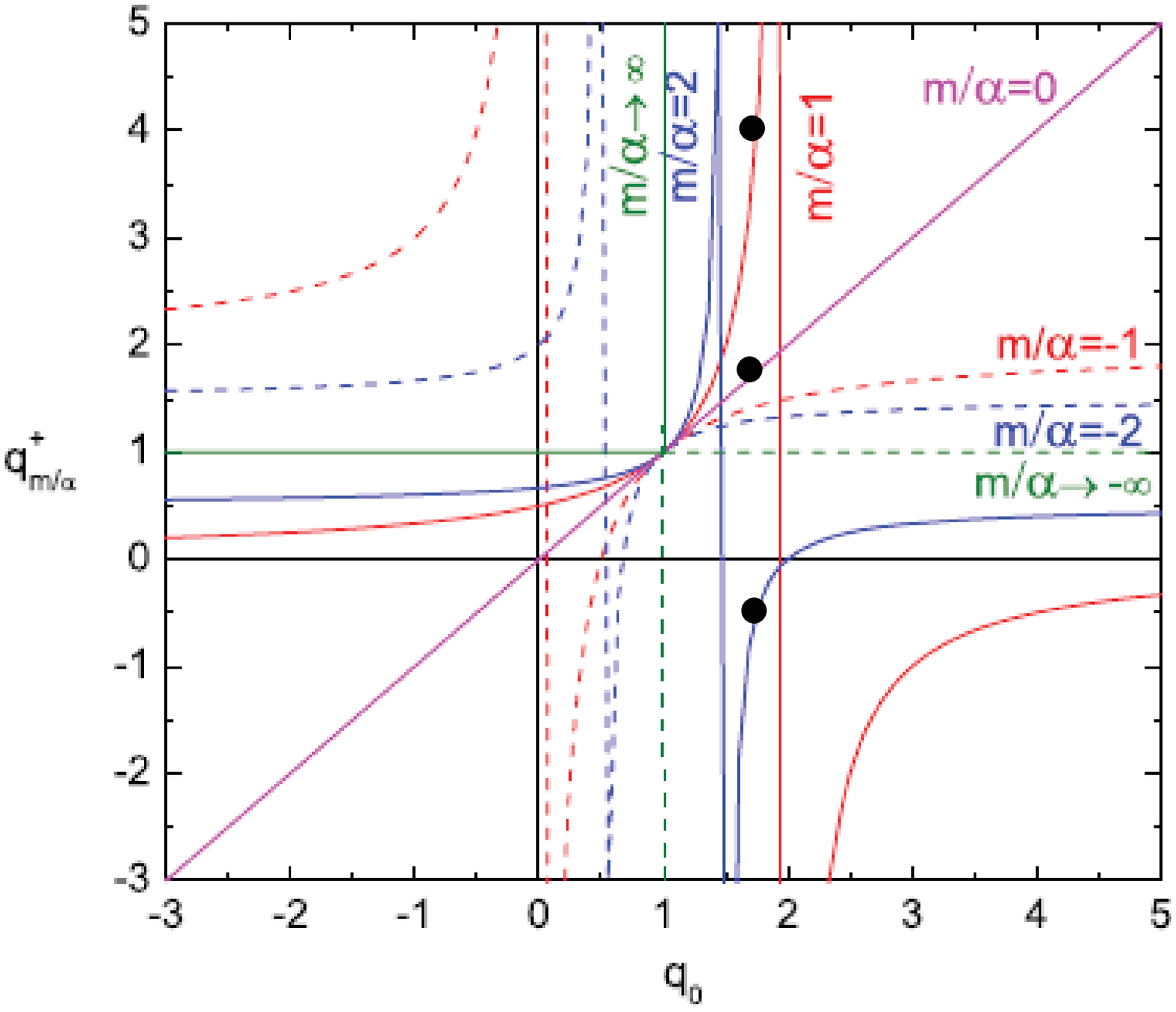,width=4.2in,angle=-90}\\
\psfig{file=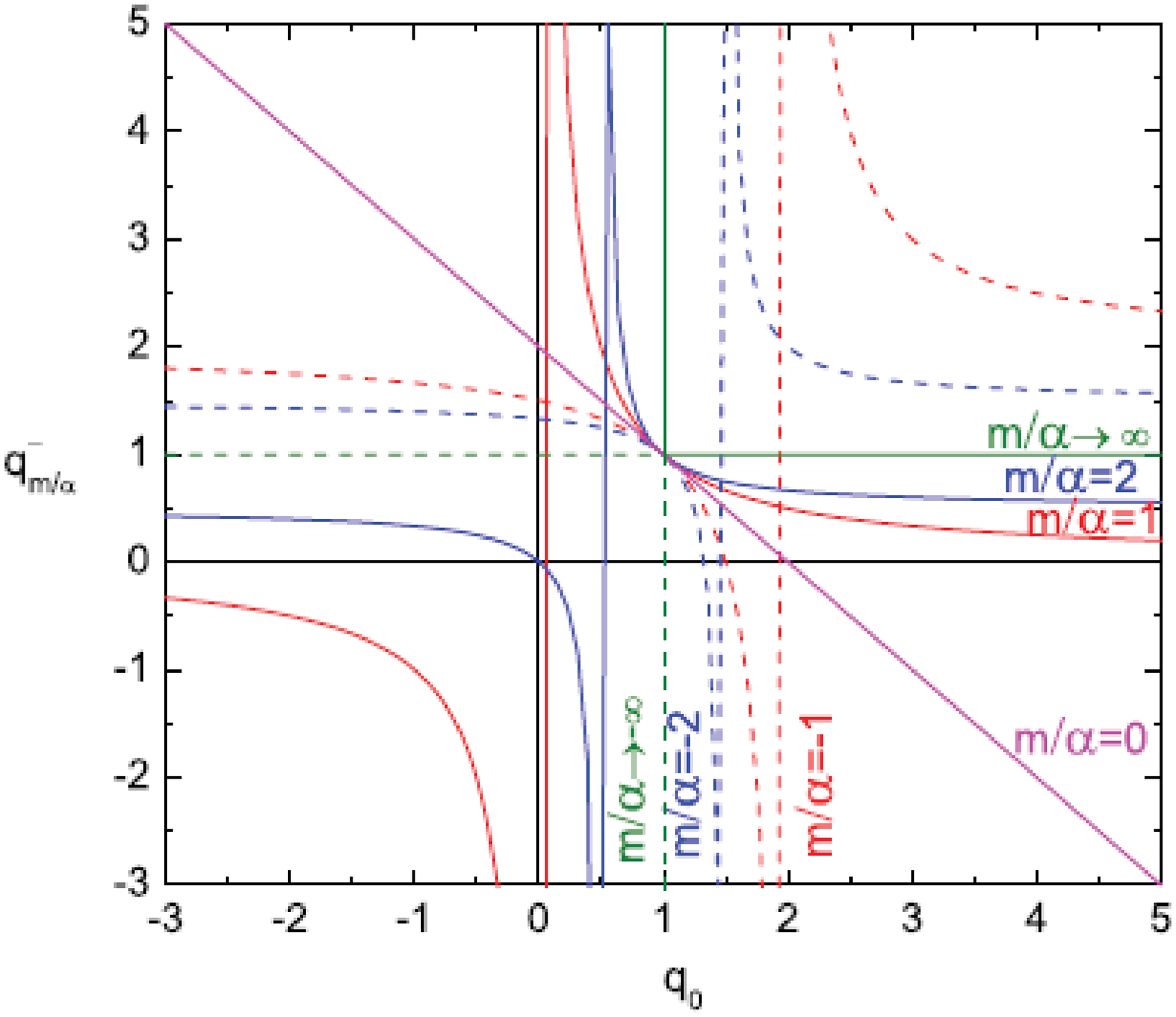,width=4in,angle=-90}
\end{center}
\caption{Curves of $q_{m/\alpha}^+$ (top) and $q_{m/\alpha}^-$ (bottom) as functions of $q_0$ for typical values of $m/\alpha$. They are respectively given by Eqs. (\ref{triplet1}) and (\ref{triplet2}). The three black dots in the top figure correspond, from top to bottom, to $q_{1}^+= q_{relaxation}=4$, $q_{0}^+= q_{stationary\,state}=7/4$ and $q_{2}^+= q_{sensitivity}=-1/2$ respectively \cite{TsallisGellMannSato2005}.}
\label{connections}
\end{figure}

Very little is known nowadays about the (plausible) analytical connections of such indices with the following algebra emerging within the $q$-CLT \cite{UmarovTsallisSteinberg2008,UmarovTsallisGellMannSteinberg2010} and related matters: 

\begin{equation}
\frac{1}{1-q_{m/\alpha}^+}= \frac{1}{1-q_0}+\frac{m}{\alpha} \;\;\;\;(0<\alpha \le 2; \, m=0, \pm1,\pm2,\pm3,...)\,,
\label{triplet1}
\end{equation}
and
\begin{equation}
\frac{1}{1-q_{m/\alpha}^-}=\frac{1}{q_0-1}+\frac{m}{\alpha} \;\;\;\;(0<\alpha \le 2; \, m=0, \pm1,\pm2,\pm3,...)\,,
\label{triplet2}
\end{equation}
where $\alpha=2$ corresponds to $q$-Gaussians, and $0<\alpha<2$ corresponds to $(q,\alpha)$-stable distributions.
See Fig. \ref{connections}. As examples of $q$-triplets and analogous relations, let us mention, among others available in the literature, the following ones:\\

(i) Fluctuations of the magnetic field of the solar wind (as detected by Voyager 1 \cite{BurlagaVinas2005}):
\begin{eqnarray}
q_{sensitivity} &=&   -0.6 \pm 0.2 \,,\nonumber      \\
q_{relaxation}  &=&   3.8 \pm 0.3 \,, \nonumber \\
q_{stationary\,state} &=& 1.75 \pm 0.06 \,, 
\end{eqnarray}
possibly interpreted \cite{TsallisGellMannSato2005} as
\begin{eqnarray}
q_{sensitivity} &=& -1/2 \,, \nonumber \\
q_{relaxation}  &=&  4  \,, \nonumber \\
q_{stationary\,state} &=& 7/4  \,.  
\end{eqnarray} \\

(ii) Edge of chaos (Feigenbaum point) of the logistic map (see \cite{TirnakliTsallisBeck2009,FuentesRobledo2010} and references therein):
\begin{eqnarray}
q_{sensitivity} &=& 0.244487701341282066198 ... \,, \nonumber \\
q_{relaxation}  &=&   2.249784109 ... \,, \nonumber \\
q_{stationary\,state} &=& 1.65 \pm 0.05  \,. 
\end{eqnarray} \\

(iii) Fluctuations of the width (above Buenos Aires) of the Ozone layer \cite{FerriReynosoPlastino2010}:
\begin{eqnarray}
q_{sensitivity} &=&   -8.1 \pm 0.2  \,, \nonumber \\
q_{relaxation}  &=&  1.32 \pm 0.06  \,, \nonumber \\
q_{stationary\,state} &=& 1.89 \pm 0.02  \,. 
\end{eqnarray} \\

As we can see, these examples suggest that quite widely $q_{sensitivity} <1<q_{stationary\,state}<q_{relaxation}$.

As open questions we may emphasize: {\it What are the precise connections between physical properties and the $q$-triplet and similar quantities? How many of those indices are independent? Are there direct connections between these indices and the $q$-CLT algebra shown in Eqs. (\ref{triplet1}) and (\ref{triplet2})?   }

\subsection{Universality classes, classification of entropies}

The BG entropy and its associated {\it exponential} distribution for thermal equilibrium have been extended, during the last two decades, in the sense of thermodynamics and statistical mechanics, into other forms. Stationary states have been discussed which correpond to the {\it $q$-exponential} form \cite{Tsallis1988}, {\it logarithmic} form \cite{Curado1999,CuradoNobre2004}, {\it stretched-exponential} form \cite{AnteneodoPlastino1999}, as well as other, more general, forms \cite{HanelThurner2011,Tempesta2010}.

The present belief is that BG entropy and statistics are sufficient but not necessary for thermodynamics. In other words, thermodynamics might be more powerful than the role attributed to it by BG statistical mechanics. A question which arises then naturally is: {\it What is the most general form of entropy which can be consistent with thermodynamics, more precisely with the zeroth, first, second, and third principles?} {\it What are the superstatistical forms} \cite{BeckCohen2003,TsallisSouza2003} {\it which would correspond to this general entropy and statistics?}

\section{Conclusions}

During the last two decades the nonadditive entropy $S_q$ and its associated nonextensive statistical mechanics have been and are being intensively studied world wide (over three thousands of papers are available in the literature \cite{Bibliography2010}). A variety of analytical, computational, experimental and observational results and applications provide nowadays what one may consider as a relatively clear understanding of its domain of validity. Nevertheless, various relevant points still remain to be clarified. We have presented here a set of such open questions in the hope that future work will improve our insights.

\nonumsection{Acknowledgments} \noindent Warm hospitality by Prof. Armin Bunde and his collaborators in Giessen University, as well as support through a Mercator Guest Professorship of the DFG-Deutsche Forschungsgemeinschaft are gratefully acknowledged. Longstanding and fruitful exchanges with H.J. Hilhorst and E.M.F. Curado are gratefully acknowledged as well. Finally, my thanks also go to the Greek team who, in Thessaloniki, made the present occasion -- warmly dedicated to our great friend and scientist Tassos Bountis -- a memorable one.


\begin{thebibliography}{9}

\bibitem[Gibbs(1902)]{Gibbs1902}Gibbs, J.W. [1902], {\it Elementary Principles in Statistical Mechanics - Developed with Especial Reference to the Rational Foundation of Thermodynamics}, C. Scribner's Sons, New York, 1902; Yale University Press, New Haven, (1981), page 35.

\bibitem[Fermi(1936)]{Fermi1936}Fermi, E. [1936], {\it Thermodynamics} (Dover). 

\bibitem[Tsallis(1988)]{Tsallis1988}Tsallis, C. [1988], ``Possible generalization of Boltzmann-Gibbs statistics",  J. Stat. Phys. {\bf 52}, 479-487. 

\bibitem[Penrose(1970)]{Penrose1970}Penrose, O. [1970], {\it Foundations of Statistical Mechanics: A Deductive Treatment}, (Pergamon, Oxford), page 167. 

\bibitem[Tsallis et al(1998)]{TsallisMendesPlastino1998}Tsallis,C., Mendes, R.S. \& Plastino, A.R. [1998], ``The role of constraints within generalized nonextensive statistics", Physica A {\bf 261}, 534-554. 

\bibitem[Curado \& Tsallis(1991)]{CuradoTsallis1991}Curado, E.M.F. \& Tsallis, C. [1991],  ``Generalized statistical mechanics: connection with thermodynamics", J. Phys. A {\bf 24}, L69-72; Corrigenda: {\bf 24}, 3187 (1991) and {\bf 25}, 1019 (1992).

\bibitem[Tsallis \& Brigatti(2004)]{TsallisBrigatti2004}C. Tsallis \& E. Brigatti [2004], ``Nonextensive statistical mechanics: A brief introduction", Continuum Mechanics and Thermodynamics {\bf 16},  223. 

\bibitem[Tsallis(2009a)]{Tsallis2009a}Tsallis, C. [2009a], {\it Introduction to Nonextensive Statistical Mechanics - Approaching a Complex World} (Springer, New York).

\bibitem[Tsallis(2009b)]{Tsallis2009b}Tsallis, C. [2009b], ``Entropy", in {\it Encyclopedia of Complexity and Systems Science}, ed. R.A. Meyers (Springer, Berlin), 11 volumes [ISBN: 978-0-387-75888-6].  

\bibitem[Tsallis(2010)]{BibliographyNEXT}Tsallis, C. [2010]. A regularly updated bibliography can be seen at http://tsallis.cat.cbpf.br/biblio.htm 

\bibitem[Upadhyaya et al(2001)]{UpadhyayaRieuGlazierSawada2001}A. Upadhyaya, J.-P. Rieu, J.A. Glazier \& Y. Sawada [2001], ``Anomalous diffusion and non-Gaussian velocity distribution of Hydra cells in cellular aggregates", Physica A {\bf 293}, 549.

\bibitem[Reynolds(2010)]{Reynolds2010}A.M. Reynolds [2010], ``Can spontaneous cell movements be modelled as L\'evy walks?", Physica A {\bf 389}, 273.

\bibitem[Daniels et al (2004)]{DanielsBeckBodenschatz2004}K.E. Daniels, C. Beck \& E. Bodenschatz [2004], ``Defect turbulence and generalized statistical mechanics", Physica D {\bf 193}, 208.

\bibitem[Douglas et al(2006)]{DouglasBergaminiRenzoni2006}P. Douglas, S. Bergamini \& F. Renzoni [2006], ``Tunable Tsallis distributions in dissipative optical lattices", Phys. Rev. Lett. {\bf 96}, 110601; G.B. Bagci \& U. Tirnakli [2009], ``Self-organization in dissipative optical lattices", Chaos {\bf 19}, 033113.

\bibitem[Arevalo et al(2007a)]{ArevaloGarcimartinMaza2007a}R. Arevalo, A. Garcimartin \& D. Maza [2007a], ``Anomalous diffusion in silo drainage", Eur. Phys. J. E {\bf 23}, 191-198.

\bibitem[Arevalo et al(2007b)]{ArevaloGarcimartinMaza2007b}R. Arevalo, A. Garcimartin \& D. Maza [2007b], ``A non-standard statistical approach to the silo discharge", in {\it Complex Systems - New Trends and Expectations}, eds. H.S. Wio, M.A. Rodriguez and L. Pesquera, Eur. Phys. J.-Special Topics {\bf 143}.

\bibitem[Liu \& Goree(2008)]{LiuGoree2008}B. Liu \& J. Goree [2008], ``Superdiffusion and non-Gaussian statistics in a driven-dissipative 2D dusty plasma", Phys. Rev. Lett. {\bf 100}, 055003.

\bibitem[DeVoe(2009)]{DeVoe2009}R.G. DeVoe [2009], ``Power-law distributions for a trapped ion interacting with a classical buffer gas", Phys. Rev. Lett. {\bf 102}, 063001.

\bibitem[Borland(2002a)]{Borland2002a}L. Borland [2002a], ``Closed form option pricing formulas based on a non-Gaussian stock price model with statistical feedback", Phys. Rev. Lett. {\bf 89}, 098701.

\bibitem[Borland(2002b)]{Borland2002b}L. Borland [2002b], ``A theory of non-gaussian option pricing", Quantitative Finance {\bf 2}, 415.

\bibitem[Osorio et al(2004)]{OsorioBorlandTsallis2004}R. Osorio, L. Borland  \& C. Tsallis [2004],  ``Distributions of high-frequency stock-market observables", in {\it Nonextensive Entropy - Interdisciplinary Applications}, eds. M. Gell-Mann and C. Tsallis (Oxford University Press, New York).

\bibitem[Queiros(2005)]{Queiros2005}S.M.D. Queiros [2005], ``On non-Gaussianity and dependence in financial in time series: A nonextensive approach", Quant. Finance {\bf 5}, 475.

\bibitem[Burlaga \& Vinas(2005)]{BurlagaVinas2005}L.F. Burlaga \& A.F.-Vinas [2005], ``Triangle for the entropic index $q$ of non-extensive statistical mechanics observed by Voyager 1 in the distant heliosphere", Physica A {\bf 356}, 375.

\bibitem[Burlaga \& Ness(2009)]{BurlagaNess2009}L.F. Burlaga \& N.F.  Ness [2009], ``Compressible ``turbulence" observed in the heliosheath by Voyager 2", Astrophys. J.  {\bf 703}, 311.

\bibitem[Bakar \& Tirnakli(2009)]{BakarTirnakli2009}B. Bakar \& U. Tirnakli [2009], ``Analysis of self-organized criticality in Ehrenfest's dog-flea model", Phys. Rev. E {\bf 79}, 040103(R).

\bibitem[Bakar \& Tirnakli(2010)]{BakarTirnakli2010}B. Bakar \& U. Tirnakli [2010], ``Return distributions in dog-flea model revisited", Physica A {\bf 389}, 3382.

\bibitem[Celikoglu et al(2010)]{CelikogluTirnakliQueiros2010}A. Celikoglu, U. Tirnakli \& S.M.D. Queiros [2010], ``Analysis of return distributions in the coherent noise model", Phys. Rev. E {\bf 82}, 021124.

\bibitem[Caruso et al(2007)]{CarusoPluchinoLatoraVinciguerraRapisarda2007}F. Caruso, A. Pluchino, V. Latora, S. Vinciguerra \& A. Rapisarda [2007], ``Analysis of self-organized criticality in the Olami-Feder-Christensen model and in real earthquakes", Phys. Rev. E {\bf 75}, 055101(R).

\bibitem[Moyano \& Anteneodo(2006)]{MoyanoAnteneodo2006}L.G. Moyano \& C. Anteneodo [2006], ``Diffusive anomalies in a long-range Hamiltonian system", Phys. Rev. E {\bf 74}, 021118.

\bibitem[Carvalho et al(2008)]{CarvalhoSilvaNascimentoMedeiros2008}J.C. Carvalho, R. Silva, J.D. do Nascimento \& J.R. de Medeiros [2008], ``Power law statistics and stellar rotational velocities in the Pleiades", Europhys. Lett. {\bf 84}, 59001.

\bibitem[Pickup et al(2009)]{PickupCywinskiPappasFaragoFouquet2009}R.M. Pickup, R. Cywinski, C. Pappas, B. Farago \& P. Fouquet [2009], ``Generalized spin glass relaxation", Phys. Rev. Lett. {\bf 102}, 097202.

\bibitem[Ferri et al(2010)]{FerriReynosoPlastino2010}G.L. Ferri, M.F. Reynoso Savio \& A. Plastino [2010], ``Tsallis' $q$-triplet and the ozone layer", Physica A {\bf 389}, 1829.

\bibitem[Bediaga et al(2000)]{BediagaCuradoMiranda2000}I. Bediaga, E.M.F. Curado \& J. Miranda [2000], ``A nonextensive thermodynamical equilibrium approach in $e^+e^- \rightarrow hadrons$", Physica A {\bf 286}, 156.

\bibitem[Wilk \& Wlodarczyk(2009)]{WilkWlodarczyk2009}G. Wilk \& Z. Wlodarczyk [2009], ``Power laws in elementary and heavy-ion collisions - A story of fluctuations and nonextensivity?", Eur. Phys. J. A {\bf 40}, 299.

\bibitem[Biro et al(2009)]{BiroPurcselUrmossy2009}T.S. Biro, G. Purcsel \& K. Urmossy [2009], ``Non-extensive approach to quark matter", in {\it Statistical Power-Law Tails in High Energy Phenomena}, Eur. Phys. J. A {\bf 40}, 325.

\bibitem[CMS1(2010)]{CMS1}V. Khachatryan et al (CMS Collaboration) [2010a], ``Transverse-momentum and pseudorapidity distributions of charged hadrons in pp collisions at $\sqrt s = 0.9$ and $2.36 \;TeV$", J. High Energy Phys. {\bf 02}, 041.

\bibitem[CMS2(2010)]{CMS2}V. Khachatryan et al (CMS Collaboration) [2010b], ``Transverse-momentum and pseudorapidity distributions of charged hadrons in $pp$ collisions at $\sqrt s = 7 \;TeV$", Phys. Rev. Lett. {\bf 105}, 022002.

\bibitem[PHENIX(2010)]{PHENIX}Adare et al (PHENIX Collaboration) [2010], ``Measurement of neutral mesons in $p + p$ collisions at $\sqrt s = 200 \; GeV$ and scaling properties of hadron production", 1005.3674 [hep-ex].

\bibitem[Shao et al(2010)]{ShaoYiTangChenLiXu2010}M. Shao, L. Yi, Z.B. Tang, H.F. Chen, C. Li \& Z.B. Xu [2010], ``Examination of the species and beam energy dependence of particle spectra using Tsallis statistics", J. Phys. G {\bf 37} (8), 085104.

\bibitem[Kaniadakis(1996)]{KaniadakisLavagnoQuarati1996}G. Kaniadakis, A. Lavagno \& P. Quarati [1996], ``Generalized statistics and solar neutrinos", Phys. Lett. B {\bf 369}, 308.

\bibitem[Tsallis et al(2003)]{TsallisAnjosBorges2003}C. Tsallis, J.C. Anjos \& E.P. Borges [2003], ``Fluxes of cosmic rays: A delicately balanced stationary state", Phys. Lett. A {\bf 310}, 372.

\bibitem[Lyra \& Tsallis(1998)]{LyraTsallis1998}M.L. Lyra \& C. Tsallis [1998], ``Nonextensivity and multifractality in low-dimensional dissipative systems", Phys. Rev. Lett. {\bf 80}, 53.

\bibitem[Borges et al(2002)]{BorgesTsallisAnanosOliveira2002}E.P. Borges, C. Tsallis, G.F.J. Ananos \& P.M.C. Oliveira [2002], ``Nonequilibrium probabilistic dynamics at the logistic map edge of chaos", Phys. Rev. Lett. {\bf 89}, 254103.

\bibitem[Ananos \& Tsallis(2004)]{AnanosTsallis2004}G.F.J. Ananos \& C. Tsallis [2004], ``Ensemble averages and nonextensivity at the edge of chaos of one-dimensional maps", Phys. Rev. Lett. {\bf 93}, 020601.

\bibitem[Baldovin \& Robledo(2004)]{BaldovinRobledo2004}F. Baldovin \& A. Robledo [2004], ``Nonextensive Pesin identity. 
Exact renormalizat-\\ion group analytical results for the dynamics at the edge of chaos of the logistic map", Phys. Rev. E {\bf 69}, 045202(R).

\bibitem[Mayoral \& Robledo (2005)]{MayoralRobledo2005}E. Mayoral \& A. Robledo [2005], ``Tsallis' $q$ index and Mori's $q$ phase transitions at edge of chaos", Phys. Rev. E {\bf 72}, 026209.

\bibitem[Pluchino et al(2007)]{PluchinoRapisardaTsallis2007}A. Pluchino, A. Rapisarda \& C. Tsallis [2007], ``Nonergodicity and central limit behavior in long-range Hamiltonians", Europhys. Lett. {\bf 80}, 26002.

\bibitem[Pluchino et al(2008)]{PluchinoRapisardaTsallis2008}A. Pluchino, A. Rapisarda \& C. Tsallis [2008], ``A closer look at the indications of $q$-generalized Central Limit Theorem behavior in quasi-stationary states of the HMF model", Physica A {\bf 387}, 3121.

\bibitem[Miritello et al(2009)]{MiritelloPluchinoRapisarda2009}G. Miritello, A. Pluchino \& A. Rapisarda [2009], ``Central limit behavior in the Kuramoto model at the 'edge of chaos' ", Physica A {\bf 388}, 4818.

\bibitem[Leo et al(2010)]{LeoLeoTempesta2010}M. Leo, R.A. Leo \& P. Tempesta [2010], ``Thermostatistics in the neighborhood of the $\pi$-mode solution for the Fermi-Pasta-Ulam $\beta$ system: From weak to strong chaos", J. Stat. Mech. P04021.

\bibitem[White et al(2006)]{WhiteKejzarTsallisFarmerWhite2006}D.R. White, N. Kejzar, C. Tsallis,D. Farmer \& S. White [2006], ``A generative model for feedback networks", Phys. Rev. E {\bf 73}, 016119.

\bibitem[Thurner et al(2007)]{ThurnerKyriakopoulosTsallis2007}S. Thurner, F. Kyriakopoulos \& C. Tsallis [2007], ``Unified model for network dynamics exhibiting nonextensive statistics", Phys. Rev. E {\bf 76}, 036111.

\bibitem[Sotolongo et al(2010)]{Sotolongo-GrauRodriguez-PerezAntoranzSotolongo-Costa2010}O. Sotolongo-Grau, D. Rodriguez-Perez, J.C. Antoranz \& O. Sotolongo-Costa [2010], ``Tissue radiation response with maximum Tsallis entropy", Phys. Rev. Lett. {\bf 105}, 158105 (4 pages).

\bibitem[Andrade et al(2010)]{AndradeSilvaMoreiraNobreCurado2010}J. S. Andrade Jr., G.F.T. da Silva, A.A. Moreira, F.D. Nobre and E.M.F. Curado [2010], ``Thermostatistics of overdamped motion of interacting particles", Phys. Rev. Lett. {\bf 105}, 260601.

\bibitem[Tamarit et al(1998)]{TamaritCannasTsallis1998}F.A. Tamarit, S.A. Cannas \& C. Tsallis [1998], ``Sensitivity to initial conditions in the Bak-Sneppen model of biological evolution", Eur. Phys. J. B {\bf 1}, 545.   

\bibitem[Anteneodo \& Tsallis(1997)]{AnteneodoTsallis1997}C. Anteneodo \& C. Tsallis [1997], ``Two-dimensional turbulence in pure-electron plasma: A nonextensive thermostatistical description",  J.  Molecular Liquids {\bf 71}, 255.  

\bibitem[Tsallis et al(2005)]{TsallisGellMannSato2005}C. Tsallis, M. Gell-Mann \& Y. Sato [2005], ``Asymptotically scale-invariant occupancy of phase space makes the entropy $S_q$ extensive", Proc. Natl. Acad. Sc. USA {\bf 102}, 15377-15382.

\bibitem[Tsallis et al(1997)]{TsallisPlastinoZheng1997}C. Tsallis, A.R. Plastino and W.-M. Zheng [1997], ``Power-law sensitivity to initial conditions - New entropic representation", Chaos, Solitons and Fractals {\bf 8}, 885-891.

\bibitem[Baldovin and Robledo(2002a)]{BaldovinRobledo2002a}F. Baldovin and A. Robledo [2002a], ``Sensitivity to initial conditions at bifurcations in one-dimensional nonlinear maps: Rigorous nonextensive solutions", Europhys. Lett. {\bf 60}, 518.

\bibitem[Baldovin and Robledo(2002b)]{BaldovinRobledo2002b}F. Baldovin and A. Robledo [2002b], ``Universal renormalization-group dynamics at the onset of chaos in logistic maps and nonextensive statistical mechanics", Phys. Rev. E {\bf 66}, R045104.

\bibitem[Robledo(2006)]{Robledo2006}A. Robledo [2006], ``Incidence of nonextensive thermodynamics in temporal scaling at Feigenbaum points", Physica A {\bf 370}, 449-460. 

\bibitem[Grassberger and Scheunert(1981)]{GrassbergerScheunert1981}P. Grassberger and M. Scheunert [1981], ``Some more universal scaling laws for critical mappings", J. Stat. Phys. {\bf 26}, 697.

\bibitem[Schneider et al(1987)]{SchneiderPolitiWurtz1987}T. Schneider, A. Politi and D. Wurtz [1987], ``Resistance and eigenstates in a tight-binding model with quasi-periodic potential", Z. Phys. B {\bf 66}, 469.

\bibitem[Anania \& Politi(1988)]{AnaniaPoliti1988}G. Anania and A. Politi [1988], ``Dynamical behavior at the onset of chaos", Europhys. Lett. {\bf 7}, 119.

\bibitem[Hata et al(1989)]{HataHoritaMori1989}H. Hata, T. Horita and H. Mori [1989], ``Dynamic description of the critical $2^{\infty}$ attractor and $2^m$-band chaos", Progr. Theor. Phys. {\bf 82}, 897.

\bibitem[Mori et al(1989)]{MoriHataHoritaKobayashi1989}H. Mori, H. Hata, T. Horita and T. Kobayashi [1989], ``Statistical mechanics of dynamical systems", Progr. Theor. Phys. Suppl. {\bf 99}, 1.

\bibitem[Latora et al(2000)]{LatoraBarangerRapisardaTsallis2000}V. Latora, M. Baranger, A. Rapisarda and C. Tsallis [2000], ``The rate of entropy increase at the edge of chaos", Phys. Lett. A {\bf 273}, 97.

\bibitem[Caruso \& Tsallis(2008)]{CarusoTsallis2008}F. Caruso and C. Tsallis [2008], ``Nonadditive entropy reconciles the area law in quantum systems with classical thermodynamics", Phys. Rev. E {\bf 78}, 021101.

\bibitem[Saguia \& Sarandy(2010)]{SaguiaSarandy2010}A. Saguia and M.S. Sarandy [2010], ``Nonadditive entropy for random quantum spin-$S$ chains", Phys. Lett. A {\bf 374}, 3384-3388.

\bibitem[Nivanen et al(2003)]{NivanenLeMehauteWang2003}L. Nivanen, A. Le Mehaute \& Q.A. Wang [2003], ``Generalized algebra within a nonextensive statistics", Rep. Math. Phys. {\bf 52}, 437. 

\bibitem[Borges(2004)]{Borges2004}E.P. Borges [2004], ``A possible deformed algebra and calculus inspired in nonextensive thermostatistics", Physica A {\bf 340}, 95.

\bibitem[Tsallis \& Queiros(2007)]{TsallisQueiros2007}C. Tsallis \& S.M.D. Queiros [2007], ``Nonextensive statistical mechanics and central limit theorems I - Convolution of independent random variables and $q$-product", in  {\it Complexity, Metastability and Nonextensivity},  eds. S. Abe, H.J. Herrmann, P. Quarati, A. Rapisarda and C. Tsallis, American Institute of Physics Conference Proceedings {\bf 965}, 8-20 (New York).

\bibitem[Queiros \& Tsallis(2007)]{QueirosTsallis2007}S.M.D. Queiros \& C. Tsallis [2007], ``Nonextensive statistical mechanics and central limit theorems II - Convolution of     $q$-independent random variables", in  {\it Complexity, Metastability and Nonextensivity},  eds. S. Abe, H.J. Herrmann, P. Quarati, A. Rapisarda and C. Tsallis, American Institute of Physics Conference Proceedings {\bf 965}, 21-33 (New York).

\bibitem[Curado et al(2008)]{CuradoRoditiTsallis2008}E.M.F. Curado, I. Roditi \& C. Tsallis [2008],  unpublished.

\bibitem[Umarov et al(2008)]{UmarovTsallisSteinberg2008}S. Umarov, C. Tsallis \& S. Steinberg [2008], ``On a $q$-central limit theorem consistent with nonextensive statistical mechanics", Milan J. Math. {\bf 76}, 307-328.

\bibitem[Umarov et al(2010)]{UmarovTsallisGellMannSteinberg2010}S. Umarov, C. Tsallis, M. Gell-Mann \& S. Steinberg [2010],  ``Generalization of symmetric $\alpha$-stable L\'evy distributions for $q > 1$", J. Math. Phys. {\bf 51}, 033502 (23 pages). 
 
\bibitem[Hilhorst(2009)]{Hilhorst2009}H.J. Hilhorst [2009], ``Central limit theorems for correlated variables: some critical remarks", Braz. J. Phys. {\bf 39}, 371.

\bibitem[Hilhorst(2010)]{Hilhorst2010}H.J. Hilhorst [2010], ``Note on a $q$-modified central limit theorem", J. Stat. Mech., P10023.

\bibitem[Jauregui \& Tsallis(2010a)]{JaureguiTsallis2010a}M. Jauregui \& C. Tsallis [2010a], ``On the role of $q$-Fourier transform on physical complexity", 1010.6275 [cond-mat.stat-mech].

\bibitem[Tsallis et al(1999)]{TsallisBemskiMendes1999}C. Tsallis, G. Bemski \& R.S. Mendes [1999], ``Is re-association in folded proteins a case of nonextensivity?", Phys. Lett. A {\bf 257}, 93.

\bibitem[Jauregui \& Tsallis(2010b)]{JaureguiTsallis2010b}M. Jauregui \& C. Tsallis [2010b], ``New representations of $\pi$ and Dirac delta using the nonextensive-statistical-mechanics $q$-exponential function", J. Math. Phys. {\bf 51}, 063304.

\bibitem[Nobre et al(2010)]{NobreMonteiroTsallis2010}F.D. Nobre, M.A. Rego-Monteiro \& C. Tsallis [2010], ``Nonlinear generalizations of relativistic and quantum equations with a common type of solution", preprint (2010).

\bibitem[Umarov \& Tsallis(2010)]{UmarovTsallis2010}S. Umarov \& C. Tsallis [2010], ``Limit distribution in the $q$-CLT for $q \ge 1$ can not have a compact support", 1012.1814 [cond-mat.stat-mech].

\bibitem[Vignat \& Plastino(2007)]{VignatPlastino2007}C. Vignat \& A. Plastino [2007], ``Central limit theorem and deformed exponentials", J. Phys. A {\bf 40}, F969.

\bibitem[Hahn et al(2010)]{HahnJiangUmarov2010}M.G. Hahn, X.X. Jiang \& S. Umarov [2010], ``On $q$-Gaussians and exchangeability", J. Phys. A {\bf 43} (16), 165208. 

\bibitem[Nelson \& Umarov(2008)]{NelsonUmarov2008}K.P. Nelson and S. Umarov [2008], ``The relationship between Tsallis statistics, the Fourier transform, and nonlinear coupling", 0811.3777 [cs.IT].

\bibitem[Nelson \& Umarov(2010)]{NelsonUmarov2010}K.P. Nelson  and S. Umarov [2010], ``Nonlinear statistical coupling", Physica A {\bf 389}, 2157-2163.

\bibitem[Moyano et al(2006)]{MoyanoTsallisGellMann2006}L.G. Moyano, C. Tsallis \& M. Gell-Mann [2006], ``Numerical indications of a $q$-generalised central limit theorem", Europhys. Lett. {\bf 73}, (2006).

\bibitem[Thistleton et al(2009)]{ThistletonMarshNelsonTsallis2009}W.J. Thistleton, J.A. Marsh, K.P. Nelson \& C. Tsallis [2009], ``$q$-Gaussian approximants mimic non-extensive statistical-mechanical expectation for many-body probabilistic model with long-range correlations",  Cent. Eur. J. Phys. {\bf 7}, 387.

\bibitem[Hilhorst \& Schehr(2007)]{HilhorstSchehr2007}H.J. Hilhorst \& G. Schehr [2007], ``A note on $q$-Gaussians and non-Gaussians in statistical mechanics", 	J. Stat. Mech. P06003.

\bibitem[Rodriguez et al(2008)]{RodriguezSchwammleTsallis2008}A. Rodriguez, V. Schwammle \& C. Tsallis [2008], ``Strictly and asymptotically scale-invariant probabilistic models of $N$ correlated binary random variables having {\em q}--Gaussians as $N\to \infty$ limiting distributions", J. Stat. Mech. P09006.

\bibitem[Hanel et al(2009)]{HanelThurnerTsallis2009}R. Hanel, S. Thurner \& C. Tsallis [2009], ``Limit distributions of scale-invariant probabilistic models of correlated random variables with the $q$-Gaussian as an explicit example", Eur. Phys. J. B {\bf 72}, 263-268 (2009).

\bibitem[Dirac(1928)]{Dirac1928}P.M.A. Dirac [1928], Proc. Roy. Soc. (London) A {\bf 117}, 610.

\bibitem[Liboff(2003)]{Liboff2003}R.L. Liboff [2003], {\it Introductory Quantum Mehanics}, Fourth Edition 
(Addison Wesley, San Francisco).

\bibitem[Anteneodo \& Tsallis(1998)]{AnteneodoTsallis1998}C. Anteneodo \& C. Tsallis [1998], ``Breakdown of the exponential sensitivity to the initial conditions: Role of the range of the interaction", Phys. Rev. Lett. {\bf 80}, 5313-5316.

\bibitem[Nobre \& Tsallis(1995)]{NobreTsallis1995}F.D. Nobre \& C. Tsallis [1995], ``Infinite-range Ising ferromagnet: thermodynamic limit within generalized statistical mechanics", Physica A {\bf 213}, 337; Erratum: {\bf 216}, 369 (1995).

\bibitem[Caride et al(1983)]{CarideTsallisZanette1983}A.O. Caride, C. Tsallis \& S. I. Zanette [1983],	``Criticality of the anisotropic quantum Heisenberg model on a self-dual hierarchical lattice", Phys. Rev. Lett. {\bf 51}, 145; {\bf 51}, 616 (1983).

\bibitem[Campa et al(2001)]{CampaGiansantiMoroniTsallis2001}A. Campa, A. Giansanti, D. Moroni \& C. Tsallis [2001], ``Classical spin systems with long-range interactions: Universal reduction of mixing", Phys. Lett. A {\bf 286}, 251.

\bibitem[Antoni \& Ruffo(1995)]{AntoniRuffo1995}M. Antoni \& S. Ruffo [1995], ``Clustering and relaxation in Hamiltonian long-range dynamics", Phys. Rev. E {\bf 52}, 2361-2374.

\bibitem[Albuquerque et al(2008)]{Albuquerque2008}M.P. de Albuquerque et al [2008], unpublished.

\bibitem[Beck \& Schlogl(1993)]{BeckSchlogl1993}C. Beck \& F. Schlogl [1993], {\it Thermodynamics of Chaotic Systems} (Cambridge University Press, Cambridge).

\bibitem[Tsallis(2004)]{Tsallis2004a}C. Tsallis [2004], ``What should a statistical mechanics satisfy to reflect nature?", Physica D {\bf 193}, 3.

\bibitem[Ferri et al(2005)]{FerriMartinezPlastino2005}G.L. Ferri, S. Martinez \& A. Plastino [2005], ``Equivalence of the four versions of Tsallis' statistics", J. Stat. Mech. P04009.

\bibitem[Tirnakli et al(2010)]{TirnakliTsallisBeck2009}U. Tirnakli, C. Tsallis \& C. Beck [2009], ``A closer look at time averages of the logistic map at the edge of chaos", Phys. Rev. E {\bf 79}, 056209.

\bibitem[Fuentes \& Robledo(2010)]{FuentesRobledo2010}M.A. Fuentes \& A. Robledo [2010], ``Stationary distributions of sums of marginally chaotic variables as renormalization group fixed points", J. Phys. C Series {\bf 201}, 012002, and ``Renormalization group structure for sums of variables generated by incipiently chaotic maps", J. Stat. Mech. P01001.

\bibitem[Curado(1999)]{Curado1999}E.M.F. Curado [1999], ``General aspects of the thermodynamical formalism", in {\it Nonextensive Statistical Mechanics and Thermodynamics}, eds. S.R.A. Salinas and C. Tsallis, Braz. J. Phys. {\bf 29}, 36. 

\bibitem[Curado \& Nobre(2004)]{CuradoNobre2004}E.M.F. Curado \& F.D. Nobre [2004], ``On the stability of analytic entropic forms", Physica A {\bf 335}, 94.

\bibitem[Anteneodo \& Plastino(1999)]{AnteneodoPlastino1999}C. Anteneodo \& A.R. Plastino [1999], ``Maximum entropy approach to stretched exponential probability distributions", J. Phys. A {\bf 32}, 1089.

\bibitem[Hanel \& Thurner(2011)]{HanelThurner2011}R. Hanel \& S. Thurner [2011], ``A comprehensive classification of complex statistical systems and an axiomatic derivation of their entropy and distribution functions", Europhys. Lett. {\bf 93}, 20006 (2011).

\bibitem[Tempesta(2010)]{Tempesta2010}P. Tempesta [2011], ``Group entropies", preprint.

\bibitem[Beck \& Cohen(2003)]{BeckCohen2003}C. Beck \& E.G.D. Cohen [2003], ``Superstatistics", Physica A {\bf 322}, 267.

\bibitem[Tsallis \& Souza(2003)]{TsallisSouza2003}C. Tsallis \& A.M.C. Souza [2003], ``Constructing a statistical mechanics for Beck-Cohen superstatistics", Phys. Rev. E {\bf 67}, 026106.

\bibitem[Bibliography(2010)]{Bibliography2010}Bibliography [2010] in http://tsallis.cat.cbpf.br/biblio.htm


\end{thebibliography}
\end{document}